\documentclass[12pt]{article}

\usepackage{amsmath}
\usepackage{graphicx}
\usepackage{enumerate}
\usepackage{natbib}
\usepackage{url} 

\newcommand{\blind}{1}

\usepackage{caption}
\usepackage{subcaption}
\usepackage[toc,page]{appendix}
\usepackage{pdfpages}
\usepackage{epstopdf}
\usepackage{booktabs}
\usepackage{algorithm}  
\usepackage{algorithmic}
\usepackage{enumitem}
\usepackage{multirow}
\usepackage{adjustbox}
\usepackage{enumitem}

\usepackage{amsthm,amsmath,natbib,mathtools,amssymb}
\RequirePackage[colorlinks,citecolor=blue,urlcolor=blue]{hyperref}

\newtheorem{lemma}{Lemma}
\newtheorem{theorem}{Theorem}
\newtheorem{definition}{Definition}

\newtheorem{proposition}{Proposition}

\DeclareMathOperator*{\argmin}{arg\,min}
\DeclareMathOperator*{\Limsup}{Lim\,sup}

\begin{document}

\def\spacingset#1{\renewcommand{\baselinestretch}%
{#1}\small\normalsize} \spacingset{1}


\if1\blind
{
  \title{\bf A unified framework on defining depth for point process using function smoothing}
  \author{Zishen Xu, Chenran Wang, Wei Wu \\
    Department of Statistics, Florida State University\\
    Tallahassee, FL 32306-4330}
  \maketitle
} \fi

\if0\blind
{
  \bigskip
  \bigskip
  \bigskip
  \begin{center}
    {\LARGE\bf Point process depth using function smoothing}
\end{center}
  \medskip
} \fi

\bigskip

\begin{abstract}
The notion of statistical depth has been extensively studied in multivariate and functional data over the past few decades.  In contrast, the depth on temporal point process is still under-explored.  The problem is challenging because a point process has two types of randomness: 1) the number of events in a process, and 2) the distribution of these events.  Recent studies proposed depths in a weighted product of two terms, describing the above two types of randomness, respectively.  In this paper, we propose to unify these two randomnesses under one framework by a smoothing procedure. Basically, we transform the point process observations into functions using conventional kernel smoothing methods, and then adopt the well-known functional $h$-depth and its modified, center-based, version to describe the center-outward rank in the original data.  To do so, we define a proper metric on the point processes with smoothed functions.  We then propose an efficient algorithm to estimated the defined ``center''.   We further explore the mathematical properties of the newly defined depths and study asymptotics.   Simulation results show that the proposed depths can properly rank the point process observations.  Finally, we demonstrate the new method in a classification task using a real neuronal spike train dataset.      
\end{abstract}

\noindent%
{\it Keywords:}  statistical depth, point process, function smoothing,  $h$-depth, proper metric, spike trains
\vfill

\newpage
\spacingset{1.5} 

\section{Introduction}


The statistical depth is a method to indicate the centrality of a data point with respect to a data cloud. It can provide a center-outward structure that can be used to understand the empirical distribution, similar to the empirical quantiles in univariate data. The concept of depth was firstly proposed by \citep{tukey1975mathematics} to handle multivariate data. Since then the notion of depth has been widely studied by mathematicians and statisticians. Different forms of depth were proposed and analyzed, based on different usages and criteria. The types of data that those depths were mainly applied to were multivariate data and  functional data. Some well-known multivariate depths include the half-space depth \citep{tukey1975mathematics}, the convex hull peeling depth \citep{barnett1976ordering}, the Oja depth \citep{oja1983descriptive}, the simplicial depth \citep{liu1990notion}, and the Mahalanobis depth \citep{liu1993quality}. A study by \cite{zuo2000general} discussed desirable properties of multivariate depth, which include affine invariance, maximality at the center, monotonicity relative to the deepest points, and vanishing at infinity. As a generalization of multivariate data from finite dimension to infinite dimension, functional data received a lot of attentions recently and functional depths were also well studied. To name a few, \cite{cuevas2007robust} proposed the $h$-depth, which applied the Gaussian kernel and $\mathbb{L}_2$ norm to form the depth. \cite{cuesta2008random} extended the idea of half-space depth and defined the random Tukey depth. In 2009, the band depth and modified band depth were introduced by \cite{lopez2009concept}, which were very commonly used in functional data problems. Similar to the work by \cite{zuo2000general}, \cite{nieto2016topologically} examined the desirable properties of functional depth, including distance invariance, maximality at center, strictly decreasing with respect to the deepest point, upper semi-continuity, and receptivity to convex hull width across the domain and continuity.

As a special type of data, observation from an orderly temporal point process is made up by an ascending sequence of event times. Such observation contains two types of randomness: the number of events and the time locations of these events.  If we treat each observation as a vector of the event times, then the dimension of this vector will be a random variable. Given this dimension, this vector will also be random with ascending entries. 
We point out that the study on the notion of depth of point process is relatively new. 
So far the only existing work in this area was done in \cite{liu2017generalized}, where the depth structure was built in two steps: For an observation $s$, 1) estimating the probability of getting the number of events $P(|s|)$; 2) given the number of events, compute the conditional depth $D(s$ $|$ $|s|)$. The estimation of probability was done through a normalized Poisson mass function and the conditional depth adopted the Mahalanobis depth for multivariate data. The final depth was the multiplication of these two with a weight power $r$: $D(s)=P(|s|)^rD(s$ $|$ $|s|)$. This depth structure involves both types of randomness and satisfies good mathematical properties such as invariant to time-shift and linear transformation, monotone on rays, and upper semi-continuous. However, the combination of the randomness on the number of events and the event time distribution is not natural that the impact of the number of events on the final depth should be adjusted by the hyper-parameter $r$. The selection of the hyper-parameter could be tricky because an inappropriate value will make one of the two randomnesses dominates the depth value. In addition, the two-step procedure deals with the number of events and the event time independently. A more desirable method should be able to combine both steps in one framework, where the two types of randomness can be measured at the same time.


To deal with these issues, we propose another approach to process the point process observation: a ``transformation'' through smoothing method \citep{wand1994kernel}. The idea is to smooth the observed point process sequence with an appropriate kernel function. Through this operation, the event time vector will be replaced by a function curve. In Section \ref{sec: Prop Met}, we will see that the point process observation and the function curve are one-to-one matched if the kernel function used for smoothing satisfies certain mild conditions. According to this bijective relation, we are able to apply the methods on functional data to the point process observations, such as metrics on functions and functional depths. In this way, both types of randomness are taken into account under one framework. For individual event time, it determines the location of the kernel function curve.  Because we take the sum of the kernel functions in the smoothing procedure, a larger number of events will enhance the smoothed curve vertically. In addition, irrespective of the number of entries in the vector, its smoothed curve will be just one function.  That is, vectors with different dimensions can be natuarally compared. 
Therefore, this approach will not suffer the same issues in the framework proposed by \cite{liu2017generalized} and it has the advantage to utilize existing depth methods for functional data.

The rest of this manuscript is organized as follows. In section \ref{sec: Methods}, we will introduce the details of transforming a vector to a smoothed curve. 
We will also propose a proper metric for the space of point process observations, followed by a discussion of the properties for this metric. With this metric, we will continue to define new depth methods for the point process observations. 
In section \ref{sec: AsyTheory}, we will examine the asymptotic theory of the proposed depths. In section \ref{sec: AppResult}, we will apply our methods on simulations and experimental datasets to validate the effectiveness of the new method. Finally in section \ref{sec: SumFW}, a summary of the paper will be shown, followed by the future work.  All mathematical proofs and algorithmic details are given in appendices in the supplementary material.

\section{Methods}
\label{sec: Methods}

\subsection{Equivalent Representation via Function Smoothing}
Our goal is to rank point process observations on a finite interval $[0, T]$ via a kernel smoothing method. We will at first introduce the type of smoothing kernels for this purpose. 

\subsubsection{Kernel Functions}
\label{basis}
The basic idea of kernel smoothing on an orderly temporal point process is to assign a kernel to each observed point and then sum over all the assigned kernels to get a smooth function.  Often the kernel is a probability density function of a given distribution such as a Gaussian kernel.
The kernel function will depend on the time interval $[0, T]$ for the point process and we denote it as $K(\cdot; T)$.
In general, we propose to use any kernel function which satisfies the following four basic requirements.  These requirements are needed in order to achieve good mathematical properties for the new depth function:
\begin{enumerate}
\item Continuous and non-negative: $K(\cdot; T)$: $(-\infty, \infty) \rightarrow [0, \infty)$ is continuous; \label{c1}
\item Positive at zero: $K(0; T)>0$;
\item Linear independence with shifting: for any $n$ ($n \in \mathbb N$) different values $t_1< t_2< \cdots < t_n$, we have: $\sum_{i=1}^{n} \alpha_i K(x-t_i; T) \doteq 0$ for any $x \in [0,T]$ $\iff$ $\alpha_1= \cdots = \alpha_n = 0$; \label{c3}
\item Scale invariance: for any $x \in [0,T]$ and constant $\alpha>0$, $K(\alpha x; \alpha T) \equiv K(x;T)$;  \label{c4}
\end{enumerate}
We define a kernel function to be ``proper'' if it satisfies the above four conditions.  To have an example of a proper function, we can consider the following Gaussian kernel function:
\begin{equation}
K_{G}(x;T)=c_1 e^{ -\frac{c_2}{T^2}x^2 }
\label{eq:gk}
\end{equation}
where $c_1$, $c_2$ are two positive constants.  This result is given in the following lemma, where the proof is given in Appendix A. 
\begin{lemma}
The Gaussian kernel in Eqn. \eqref{eq:gk} is a proper kernel. 
\label{lem:gk}
\end{lemma}

\subsubsection{Proper Metric}
\label{sec: Prop Met}

Before moving on to the depth function, we look for a proper metric to measure the difference between two observed point processes.  Since the two point processes, in general, may have different number of events, multivariate measurements cannot be applied directly due to inconsistency in dimensions.  As smoothing functions are included in this depth, here we will use the distance on smoothed processes as the distance on point processes.  Let the time interval of the point process be $[0, T]$. We define the set of observed point process with given dimension $l>0$ as: $\Omega_l=\{ x=(x_1,x_2, \cdots, x_l) \ \in \mathbb R^l \quad | \quad 0 \leq x_1 \leq x_2 \leq \cdots \leq x_l \leq T \}$. For the case of $l=0$, there is no observed event in $[0,T]$, so $\Omega_0$ is the set of 0-length vector, i.e. $\Omega_0=\{\phi_0\}$ where $\phi_0$ is the event time vector of no event. Then $\Omega = \cup_{l=0}^{\infty} \Omega_l$ is the space of all point processes. Each process $x=(x_1,x_2, \cdots, x_l) \in \Omega_l$ with $l>0$ can be represented using a Dirac delta function in the form:
$x = \sum_{i=1}^l \delta(\cdot - x_i). $
Let $K(\cdot; T)$ be the smoothing kernel.  Then the smoothed process is a function on $[0, T]$ in the form:
\begin{equation}
f_{x}(t) = \sum_{i=1}^{l} K(t-x_i; T).
\label{eq:smooth}
\end{equation}
In general, the space of smoothed processes for $l$ events with $l>0$ is $\mathbb{F}_l=\{ f_{x}: [0,T] \to \mathbb{R} \quad | \quad f_{x}(t)=\sum_{i=1}^{l} K(t-x_i) \text{ where } x=(x_1,x_2, \cdots, x_l) \in \Omega_l \}$. For $l=0$, 
the smoothed process is $f_{\phi_{0}}(t) \equiv 0$ and $\mathbb{F}_0=\{ f_{\phi_{0}} \}$.  The space of all smoothing processes is then $\mathbb{F} = \cup_{l=0}^{\infty} \mathbb{F}_l$.
We point out that the correspondence between a point process and its smoothed version is one-to-one.  This is given in the following lemma (the proof is in Appendix B). 

\begin{lemma}
The smoothing procedure is a bijective mapping from $\Omega$ to $\mathbb{F}$. 
\label{lem:bijec}
\end{lemma}

Now we are ready to define a metric on the point process space.  This definition is based on the classical $\mathbb L^p$ norm on the smoothed processes.  This is formally given as follows:

\begin{definition} 
For any two point processes $s$, $t \in \Omega$ on $[0, T]$ and the correspondent $f_{s}, f_{t} \in \mathbb{F}$ given by Eqn. \eqref{eq:smooth}, we define a distance function $d_{K,p}$ between $s$ and $t$ as: 
\begin{equation}
d_{K,p}(s,t)=\| f_{s}-f_{t} \|_p
\label{eq:metric}
\end{equation}
where $\|.\|_p \ (p \ge 1)$ is the classical $\mathbb{L}^p$ norm on $[0, T]$.
\label{def:metric}
\end{definition}

Note that the smoothing kernel $K(\cdot; T)$ and $\mathbb{L}^p$ norm can influence the distance value. However, we can prove that the distance $d_{K,p}$ in Eqn. \eqref{eq:metric} is a proper metric on $\Omega$ for any $K$ and $p$.  This is given in the following theorem (the proof is in Appendix C). 

\begin{theorem}
If the smoothing kernel $K$ satisfies the four conditions to be proper, then the function $d_{K,p}$ in Definition \ref{def:metric} is a proper metric in the point process space $\Omega$.   That is, it satisfies non-negativity, identity of indiscernibles, symmetry, and triangle inequality. 
\label{thm:metric}
\end{theorem}

Based on Lemma \ref{lem:bijec} and Theorem \ref{thm:metric}, we know that the point process space $\Omega$ and smoothed process space $\mathbb{F}$ are one-to-one, and an $\mathbb L^p$ norm on $\mathbb F$ can be used to define a proper metric on $\Omega$.  This metric helps solve the problem in $\Omega$ where a conventional vector metric cannot be directly used as different processes may have different cardinalities.  Based on this result, we need to answer two natural questions: 
\begin{enumerate}
\item  For two point processes with the same cardinality, if the event times in the two processes are very close, will the distance $d_{K,p}$ between them be close, too?  
\item  Conversely, if the distance $d_{K,p}$ between two point processes is close, will they have the same cardinality and their events times are also close? 
\end{enumerate}

Question 1 examines if the $d_{K,p}$ distance is continuous with respect to the event times.   We claim that this is true and the conclusion is stated in Proposition \ref{prop:cont} as follows.  The detailed proof is given in Appendix D.  

\begin{proposition}
For any $k \in \mathbf N$, 
suppose $y=(y_1,y_2, \cdots, y_k)$ is an observed point process in $\Omega_k$.  Let
$x^{(n)}=(x_1^{(n)},x_2^{(n)}, \cdots, x_{k}^{(n)}), n = 1, 2, \cdots$ be a sequence of processes in $\Omega_k$. If 
 $\lim_{n\to \infty} x^{(n)}=y$, or equivalently, $ \lim_{n\to \infty} x_i^{(n)}=y_i $, $i=1,2, \cdots, k$, then for any observed point process $z=(z_1,z_2, \cdots, z_l) \in \Omega$,  
\begin{equation*}
\lim_{n\to \infty} d_{K,p}(x^{(n)},z)=d_{K,p}(y,z), \quad \textrm{and} \quad  \lim_{n\to \infty} d_{K,p}(x^{(n)},y)=0.   
\end{equation*}
\label{prop:cont}
\end{proposition}

\vspace{-36pt}

Question 2 examines the inverse continuity.  
We claim that this inverse continuity is also true and the result is stated in Proposition \ref{prop:invcont} (see Appendix E for a detailed proof). 

\begin{proposition}
For any $k \in \mathbf N$, 
suppose $y=(y_1,y_2, \cdots, y_k)$ is an observed point process in $\Omega_k$.  Let
$x^{(n)}=(x_1^{(n)},x_2^{(n)}, \cdots, x_{k_n}^{(n)})$ be a sequence of processes in $\Omega_{k_n}, n = 1, 2, \cdots$. 
 If $\lim_{n\to \infty} d_{K,p}(x^{(n)},y)=0$, then
$\lim_{n\to \infty} x^{(n)}=y$. 
That is,  
$$ 1) k_n=k \textrm{ for $n$ sufficiently large}, \quad \textrm{and} \quad 2) \lim_{n\to \infty} x_i^{(n)}=y_i, i=1,2, \cdots, k.$$
\label{prop:invcont}
\end{proposition}

\vspace{-48pt}

\subsection{Desirable Properties for Depth on Point Process}
\label{sec:des_prop}

In general, a depth definition is defined to provide a measure of centrality of a given data point within a data cloud. To measure if such goal is achieved, one often examines desirable mathematical properties corresponding to the centrality measurement. For instance, \cite{zuo2000general} proposed the desirable properties for multivariate depth. Later, \cite{nieto2016topologically} provided the desirable properties for functional depth. The commonly studied properties are (1)
linear invariance, (2) vanishing at infinity, (3) maximality at the center, and (4) monotonicity.  

%
%
%
%
%

\noindent Motivated by previous studies on desirable properties for multivariate and functional depths, we propose the following five desirable properties for the depth on point process: Suppose $D(s;P_S)$ is the depth function for observed event time vector $s$ with respect to the probability space $(\Omega, \mathcal F, P_S)$ for a random point process $S$ on interval $[0,T]$, we expect that $D$ satisfies
\begin{enumerate}[label={\bfseries\itshape P\arabic*}]
\item Continuity: $D(s;P_S)$ is continuous with respect to $s$ in $\Omega$. \label{des_prop1}

\item Linear invariance: 
For any event time vector $s=(s_1,s_2, \cdots, s_k)$ on interval $[0,T]$, we transform it to $\tilde{s}=(as_1+b,as_2+b, \cdots, as_k+b)$ on interval $[b,aT+b]$ where $a,b \in \mathbb{R}$ are two constants. Then  $D(\tilde{s};P_{\tilde{S}})=D(s;P_S)$. \label{des_prop2}

\item Vanishing at infinity: When the number of events goes to infinity, the depth should go to 0: $D(s;P_S)\to0$ as $|s|\to\infty$.\label{des_prop3}

\item Unique maximum at the center: There exists $s_c \in \Omega$, such that $D(s_c;P_S)=\max_{s\in\Omega}D(s,P_S)$.  Also, for any $x (\neq s_c) \in\Omega$, $D(x;P_S)<D(s_c;P_S)$. We refer to $s_c$ as the ``center''. \label{des_prop4}

\item Monotone decreasing from the center: For any $s_1.s_2\in\Omega$, suppose the center is $s_c$, if $d_{K,p}(s_1,s_c)<d_{K,p}(s_2,s_c)$, then we have $D(s_1,P_S)>D(s_2,P_S)$. \label{des_prop5}
\end{enumerate}

\noindent 
We point out that the ``center'' in \ref{des_prop4} was originally defined as a point with symmetry in the data cloud. For example, when dealing with zero-mean multivariate normal samples, we will consider the origin as center, and the corresponding Mahalanobis depth is uniquely maximized at this point. However, for point process data in a finite domain $[0, T]$, it is difficult to define a geometrically symmetric central point in $\Omega$. Notwithstanding, we expect the proposed depth has a unique maximum point, which results in our notion of ``center''. 

Here we examine if the above five properties are satisfied by the generalized Mahalanobis depth on point process \citep{liu2017generalized}: 1)  If two point processes are close to each other, they should have the same number of events and close event times in the time order.
 As the classical Mahalanobis depth is continuous with respect to the input vector, the generalized Mahalanobis depth on point process satisfies \ref{des_prop1}; 2) Linear transformation on the time will not change the number of events, so $P(|s|)^r$ will keep the same. Because the classical Mahalanobis depth is linear invariant, \ref{des_prop2} is also satisfied; 3) The Poisson mass function will go to 0 as the input goes to infinity, so $P(|s|) \to 0$ as $|s|\to\infty$. Thus, \ref{des_prop3} holds, too; 4) $D(s$ $|$ $|s|)$ has maximum value to be 1. However, this maximum in general can be achieved at multiple processes. For example, if the total intensity $\Lambda$ is an integer, both the population mean with dimension $\Lambda$ and the population mean with dimension $\Lambda-1$ will achieve the maximum depth. Thus, \ref{des_prop4} does not hold as the solution to the maximum is not unique; 5) As a result, \ref{des_prop5} does not hold either. In summary, the generalized Mahalanobis depth on point process only satisfies \ref{des_prop1} to \ref{des_prop3}.

Similar to the discussion about the Mahalanobis depth, we will refer to \ref{des_prop1} - \ref{des_prop5} to explore these properties for other depth functions in the following sections of this paper.

\subsection{$h$-depth on Point Process}
\label{sec:hd}

We have defined a proper metric in the point process space $\Omega$.  This metric is based on a smoothing procedure where a point process is equivalently represented by a function in $\mathbb F$.  In this section, we will exploit a commonly used functional depth, called $h$-depth, to define an ``$h$-depth'' for point processes in $\Omega$.   
 

At first, we review the notion of $h$-depth \citep{cuevas2007robust} for functional data.  Assume $X$ is a functional random variable in the probability space $(\Lambda, \mathcal{F}_\Lambda, P_\Lambda)$, where $\Lambda$ is a subset of $\mathbb L^2([0, T])$.  Let $\lVert \cdot \rVert$ denote the conventional $\mathbb L^2$ norm. For any function $z \in \Lambda$, its $h$-depth is defined as:
\begin{equation}
HD(z; P_\Lambda) = \mathbb E(G_h(\lVert z-X \rVert)), 
\label{eq:h_fd}
\end{equation}
where $G_h(\cdot) = \exp (-\frac{t^2}{2h})$ is a modified Gaussian kernel with parameter $h \ (>0)$ and $X$ is a random function on the probability space $(\Lambda, \mathcal{F}_\Lambda, P_\Lambda)$. 
To simplify notation, we use $HD(z)$ to represent $HD(z; P_\Lambda)$ when the measure $P_\Lambda$ is implicitly known. 
Given a set of i.i.d. random functions $X_1, X_2, \cdots, X_n \in \Lambda$, the sample version of the $h$-depth of $z$ is given as
\begin{equation*}
HD_n(z) =\frac{1}{n}\sum_{i=1}^n G_h(\lVert z-X_i \rVert)=\frac{1}{n}\sum_{i=1}^n \exp (-\frac{\lVert z-X_i \rVert^2}{2h}). 
\end{equation*}

One important property of the $h$-depth in Eqn. \eqref{eq:h_fd} is its continuity.  This is given in the following lemma, where the proof can be found in Appendix F. 

\begin{lemma}
If the functional random variable $X$ satisfies $\mathbb E\lVert X \rVert < \infty$, then $HD(\cdot)$ is a continuous function on $\Lambda$. 
\label{lem:hcont}
\end{lemma}

Based on the notion of $h$-depth on functional data (in Eqn. \eqref{eq:h_fd}) and the proper metric on point process (in Eqn. \eqref{eq:metric}), we can formally introduce the $h$-depth to point process observations as follows:  

\begin{definition}
Let $S$ be a random point process on $[0, T]$ in the probability space $(\Omega, \mathcal F, P_S)$. The $h$-depth of any $s \in \Omega$ is defined as:
\begin{equation}
D(s; P_S) = \mathbb E(G_h(\lVert f_s-f_S \rVert))
\label{eq:hdep}
\end{equation}
where $ f_s$ and $f_S$ are smoothed curves for $s$ and $S$ (by Eqn. $\eqref{eq:metric}$) and $G_h(t) = \exp (-\frac{t^2}{2h})$. 
\label{def:hdep}
\end{definition}

Definition \ref{def:hdep} provides the population version of the depth. In practice, we should follow the sample version of $h$-depth on the given observations, which is given below.

\begin{definition}
Let $\{S_i\}_{i=1}^{N}$ be a sample of event time vectors from a point process on $[0, T]$. The empirical $h$-depth of any $s \in \Omega$ is defined as:
\begin{equation}
\hat{D}(s; \{S_i\}_{i=1}^{N}) = \frac{1}{N}\sum_{i=1}^{N} G_h(\lVert f_s-f_{S_i} \rVert), 
\label{eq:hdep}
\end{equation}
where $ f_s$ and $f_{S_i} $ are smoothed curves for $s$ and $S_i $, $i=1,2, \cdots, N$, and $G_h(t) = \exp (-\frac{t^2}{2h})$ with parameter $h \ (>0)$. 
\label{def:hdep_emp}
\end{definition}

With the definitions for both the population and empirical versions of the $h$-depth on point process, we now examine the mathematical properties based on the discussion in section \ref{sec:des_prop}. The basic result can be summarized in the following proposition. 

\begin{proposition}
Let $D(\cdot)$ denote the $h$-depth on point process in Definition \ref{def:hdep}.  Then it satisfies the following properties: 
\begin{itemize}
\item \ref{des_prop1}: If $\mathbb E(|S|) < \infty$, then the depth $D(s;P_S)$ is continuous with respect to $s$.
\item \ref{des_prop2}: If the parameter $h$ is proportional to the interval length $T$, i.e. $h=C T$ for some constant $C>0$, then the depth $D(s;P_S)$ is invariant with respect to a linear transformation on the time interval.  
\item \ref{des_prop3}: $D(s; P_S) \rightarrow 0$ when $|s| \rightarrow \infty$.
\end{itemize}

\label{thm:hprop}
\end{proposition}
\noindent The detailed proof is given in Appendix G. Note that \ref{des_prop4} and \ref{des_prop5} are not satisfied because $h$-depth in general may not have a unique maximum point in the data cloud. 

\subsection{Modified $h$-depth on Point Process}
\label{sec:mhd}

The center has been a critical notion in statistical depths.  However, as we have pointed out in Sec. \ref{sec:hd}, the $h$-depth for a point process may not have a center.  In this section, we propose to modify the $h$-depth by including a ``center''-based process in the definition. 

\subsubsection{Definition and Properties}
Many commonly used depth functions have the ``center'' as the point with maximum depth.
Based on this idea, we propose a center-based new depth function on point process using the smoothing method.  The formal definition is given as follows:


\begin{definition}
Let $s_c$ be a given ``center'' point process in $\Omega$ on $[0, T]$.  For any $s \in \Omega$, its center-based $h$-depth is defined to be:
\begin{equation}
D(s;s_c) = \exp (- \frac{\| f_s - f_{s_c} \|^2}{2h}), 
\label{eq:hdep_cen}
\end{equation}
where $h > 0$ is a parameter. $ f_s$ and $f_{s_c}$ are smoothed processes of $s$ and $s_c$, respectively. 
\label{def:hdep_cen}
\end{definition}

\noindent {\bf Remark 1:} We point out that Definition \ref{def:hdep_cen} is a modified version of Definition \ref{def:hdep}. 
In fact, the classical $h$-depth takes the form
$
\mathbb{E}[\exp(-\frac{ \| z-X \|^2}{2h})],
$
with $X$ to be a functional random variable. If the order of the exponential function and the expectation is switched, then we have
$
\exp[\mathbb{E}(-\frac{ \| z-X \|^2}{2h})]
=\exp(-\frac{ \mathbb{E} \| z-\mathbb{E}X\|^2+ \mathbb{E}\|\mathbb{E}X-X \|^2+2 \mathbb{E}<z-\mathbb{E}X, \mathbb{E}X-X>}{2h}) 
= \exp(-\frac{ \| z-\mathbb{E}X\|^2+\mathbb{E}(\|\mathbb{E}X-X \|^2}{2h}) \propto
\exp(-\frac{\| z-\mathbb{E}X\|^2}{2h} )$.
In this way we obtain the center-based $h$-depth $D(s;s_c)$.

\noindent {\bf Remark 2:} The $\mathbb{L}^2$ norm in the definition can be generalized to $\mathbb{L}^p$, with $1\le p<\infty$. 


With a center process given in the definition, this modified $h$-depth for point process is expected to satisfy more desirable mathematical properties than the classical one.  Indeed, all the desirable properties in section \ref{sec:des_prop} are satisfied and the details are given in the following proposition.  

\begin{proposition}
Let $D(\cdot;s_c)$ denote the center-based $h$-depth on point process in Definition \ref{def:hdep_cen} and the metric in $\Omega$ is the $\mathbb{L}^2$ norm $d_{K,2}(s,t)=\| f_{s}-f_{t} \|$.  Then $D(\cdot;s_c)$ satisfies the following desirable mathematical properties in section \ref{sec:des_prop}:
\begin{itemize}
\item \ref{des_prop1}: If $|s|<\infty$, then depth $D(s;s_c)$ is continuous with respect to $s$.
\item \ref{des_prop2}: If the parameter $h$ is proportional to the interval length $T$, i.e. $h=CT$ for some constant $C>0$, then the depth $D(s;s_c)$ is invariant with respect to a linear transformation on the time interval. 
\item \ref{des_prop3}: $D(s;s_c) \rightarrow 0$ when $|s| \rightarrow \infty$.
\item \ref{des_prop4}:  $D(s_c;s_c) = \max_{s\in \Omega} D(s;s_c)$ and $\forall t(\neq s_c) \in\Omega$, $D(t;s_c)< D(s_c;s_c)$. 
\item \ref{des_prop5}: For any $s_1,s_2 \in \Omega$, if $d_{K,2}(s_1,s_c)<d_{K,2}(s_2,s_c)$, then $D(s_1;s_c)>D(s_2;s_c)$.
\end{itemize} 
\label{prop:hprop_cen}
\end{proposition}
\vspace{-12pt}
\noindent The detailed proof for each property is provided in Appendix H.  Note that this is a clear advantage over the classical $h$-depth method, where only properties \ref{des_prop1}-\ref{des_prop3} are satisfied. 


\subsubsection{Estimation of the Center}
\label{center_estimation}

In practice, there is no prior knowledge of the center of point process, so a proper estimation of the center will be needed. In this study, we adopt the common notion of the Karcher mean in a metric space \citep{grove1973conjugatec}.  Since we have a well-defined metric between any two point processes in Definition \ref{def:metric}, the Karcher mean can be defined as:
\begin{definition} 
Let $S$ be a random point process on $[0, T]$ in the probability space $(\Omega, \mathcal F, P_S)$.  The Karcher mean of $S$ is defined to be 
\begin{equation}
\mu_{K} = \argmin_{t \in \Omega} \mathbb E[d_{K,2}^2(t,S)],
\end{equation}
where $d_{K,2}(\cdot, \cdot)$ is the metric with kernel function $K(\cdot;T)$ and $\mathbb{L}^2$ norm in Definition \ref{def:metric}. 
\label{def:karcher}
\end{definition}
\noindent Note that, in general the solution of the minimization may not be unique, so $\mu_{K}$ is a set of points. Based on the Karcher mean in Definition \ref{def:karcher}, its empirical version is given as follows:
\begin{definition} 
Let $\{S_i\}_{i=1}^{N}$ be a collection of $N$ observed point processes on $[0, T]$. The empirical Karcher mean of this process, based on $\{S_i\}_{i=1}^{N}$, is defined to be 
\begin{equation}
\bar{S}_K^{(N)} = \argmin_{t \in \Omega} \frac{1}{N} \sum_{i=1}^{N} d_{K,2}^2(t,S_i)
\end{equation}
\label{def:ekarcher}
\end{definition}
\vspace{-24pt}
\noindent The empirical Karcher mean also may not be unique, and therefore $\bar{S}_K^{(N)}$ is a set of solution points, too. For simplicity, we define the following notation:
\begin{equation*}
SSD(t;S_1,S_2, \cdots, S_N)=\sum_{i=1}^{N} \| f_t - f_{S_i} \|_2^2
\end{equation*}
where $SSD$ stands for ``Sum of Squared Distance'' and $t\in\Omega$ is the event time vector. With this definition, the empirical Karcher mean is just $\bar{S}_K^{(N)}=\argmin_{t \in \Omega} SSD(t;S_1,S_2, \cdots, S_N)$. For any element in $\bar{S}_K^{(N)}$, its dimension has a finite upper bound. The result is formally given as follows, where the proof is given in Appendix I.
\begin{theorem}
Any solution in the empirical Karcher mean of a point process has an upper bound in dimension, i.e. $\exists D_K^{(N)}\in\mathbb{N}^+$ such that $\dim{(\bar{S})}\leq D_K^{(N)},$ $\forall \bar{S} \in \bar{S}_K^{(N)}$.
\label{KMB}
\end{theorem}
\noindent  We will introduce a theoretically-proven convergent algorithm to estimate $\bar{S}_K^{(N)}$ where this upper bound $D_K^{(N)}$ provides an effective searching range.  Such algorithm is a combination of simulation annealing and line search. We will at first briefly review the simulation annealing and line search, and then introduce an approach to combine these two methods. 



\begin{itemize}[leftmargin=*]
\item RJMCMC annealing\\
Simulated annealing \citep{geman1984stochastic, van1987simulated} is a method that makes optimization feasible if we are able to generate samples. To find the global minimum of function $f(x)$, we can build a density function $\pi_i(x) \propto \exp{[-\frac{1}{T_i}f(x)]}$ where $T_i$ is a pre-determined decreasing sequence of positive numbers called temperature and $\lim_{i\to\infty} T_i=0$ (e.g. $ T_i=C / \log(1+i) $ for constant $C>0$). Under weak regularity assumptions on $f$, the density $\pi_i$ will be concentrated on the global minimum points of $f$. After a large number of iterations of computing $\pi_i$ and sampling from it, we will have samples passing the global minimum point. 

Simulation annealing provides a method to find the minimum of a function, but to apply it on the optimization of $SSD$, a tool that is able to generate vectors with different dimensions will be needed. In this paper, we adopt the reversible-jump Markov Chain Monte Carlo (RJMCMC) method, which was first introduced by \cite{green1995reversible}. Its main idea is the Metropolis Hasting method \citep{metropolis1953equation}, but generalizes to allow the candidate to have different dimensions. In this way, it can be used to get samples from densities where the dimension of the random vector is not fixed. One condition to use the RJMCMC method is that the number of possible dimensions should be finite. According to Theorem \ref{KMB}, the range of dimension is $1, \cdots, D_K^{(N)}$, so RJMCMC is feasible for the center estimation.

Combining the two methods, we are able to find the Karcher mean through sampling from $\pi_i(t) \propto \exp{[-\frac{1}{T_i} SSD(t;S_1,S_2, \cdots, S_N)]}$. Because of RJMCMC, the annealing process can search over different dimensions automatically. An example RJMCMC annealing algorithm is provided in Appendix J, where the input to the algorithm can be adjusted before application.

\item Line search \\
According to Theorem \ref{KMB}, because the dimensions to search is finite $1,2\cdots D_K^{(N)}$, optimizing within each dimension and comparing the output across dimensions will be another feasible approach, which corresponds to the 
 ``line search'' idea. An advantage of this method is that the gradient of the $SSD$ can be computed explicitly with the input dimension fixed, as shown in the following proposition.

\begin{proposition}
If the dimension of the input event time vector $t$ is given and the distance function is $d_{k,2}$, then the gradient of $SSD$ with respect to $t$ will be:
\begin{equation*}
\frac{\partial SSD}{\partial t}(t; S_1,S_2, \cdots, S_N)=-2\int_{0}^{T} K'(x-t)[N \vec{1}_{|t|}^T  K(x-t;T) - \sum_{i=1}^{N} \vec{1}_{|S_i|}^T K(x-S_i;T) ] dx
\end{equation*}
where $t$ and $S_i$ are event time vectors. $x$ is scaler and $|t|$ is the dimension of vector $t$. All operations are done element-wise. 

\noindent If the modified Gaussian kernel $K_G(t;T)=c_1 e^{-\frac{c_2}{T^2}t^2}$ is used to smooth the process, then the gradient can be further simplified to:
\begin{equation*}
\frac{\partial SSD}{\partial t}(t; S_1,S_2, \cdots, S_N)
=
\frac{4c_1^2c_2}{T^2}[N g(t \vec{1}_{|t|}^T,\vec{1}_{|t|} t^T) \vec{1}_{|t|}- 
\sum_{i=1}^{N} g(t \vec{1}_{|t|}^T, \vec{1}_{|S_i|} S_i^T)\vec{1}_{|S_i|} ] 
\end{equation*}
with the function $g(\cdot,\cdot)$ to be
\begin{eqnarray*}
g(x,y)& = &e^{-\frac{c_2}{2T^2}(x-y)^2} \{ \frac{T^2}{4c_2}[ e^{-\frac{2c_2}{T^2}(\frac{x+y}{2})^2}  - e^{-\frac{2c_2}{T^2}(T-\frac{x+y}{2})^2}] \\
& &- \sqrt{\frac{\pi}{8c_2}}T(x-y) [ \Phi(\frac{2\sqrt{c_2}}{T}(T-\frac{x+y}{2}))-\Phi(-\frac{\sqrt{c_2}(x+y)}{T}) ] \}
\end{eqnarray*}
where $\Phi$ is the cumulative distribution function for standard Normal distribution $N(0,1)$; $\vec{1}_d=[1,1, \cdots, 1]^T\in\mathbb{R}^d$. All operations are done element-wise. 
\label{prop:gradient}
\end{proposition}

The proof and computational details are provided in Appendix L. Due to this property, gradient-based methods, such as stochastic gradient descent, can be applied to efficiently do the optimization within a given dimension. Then a comparison of the output will return the solution to the minimum $SSD$, which is the empirical Karcher mean. An example line search algorithm based on gradient method is provided in Appendix K.

\item Combined method\\
Both RJMCMC annealing method and line search method can be used to estimate the empirical Karcher mean, but they have some disadvantages. For the RJMCMC annealing, although it can search over dimensions automatically, it converges slowly to the optimal solution. 
For line search, it can converge fast to the solution given the dimension, but it needs to search every dimension in the dimension range. As a result, the time cost for the entire line search algorithm is also very large.

Based on their characteristics, if we combine the two methods: do RJMCMC annealing first as a ``pre-train'' process to locate a range of optimal dimensions and then use line search to do optimization within the narrowed dimension range, the computation efficiency is expected to be significantly improved. Moreover, the output event time vectors from the RJMCMC annealing can be used as the initial values for the line search. In this way, the number of iterations in the line search will be reduced and it can also help avoid local optimal points. The combined algorithm following this idea is shown in Algorithm \ref{alg:comb}.  The application of this algorithm on simulations and real data will be shown in section \ref{sec: AppResult}.
\end{itemize}

\begin{algorithm}[!ht]
\caption{Combined method to find empirical Karcher mean}
\begin{algorithmic}
\label{alg:comb}
\STATE {\textbf{Input}: the observed event time vectors $S_1, S_2, \cdots, S_N$. \\
\textit{For RJMCMC annealing}: The maximum number of iterations $n_{max}$; Initial value $x_0$; Initial dimension $k_0$ to be the dimension of $x_0$; the upper bound of dimension $D_K^{(N)}$. \\
\textit{For line search}: The batch size $B$; the learning rate $r$; the maximum number of epochs $ep_{max}$; convergence indicator $\epsilon$. \\
\textit{For combination}: The number of dimensions to be kept in RJMCMC annealing: $d_r\in \mathbb{N}$. }

\STATE { \textbf{(1) Pre-train:} \\
Use RJMCMC annealing (such as Algorithm 2) with input $S_1, S_2, \cdots, S_N$, $n_{max}$, $x_0$, $k_0$ and $D_K^{(N)}$ to find the top $d_r$ number of events $\{k_{0,i}\}_{i=1}^{d_r}$ and the corresponding event time vectors $\{x_{0,i}\}_{i=1}^{d_r}$ that produce the smallest $SSD$ values.}

\STATE { \textbf{(2) Optimization:} \\ 
Use line search (such as Algorithm 3) with input $S_1, S_2, \cdots, S_N$, $B$, $r$, $ep_{max}$, $\epsilon$ as parameters, and the output from RJMCMC annealing: $\{k_{0,i}\}_{i=1}^{d_r}$ and $\{x_{0,i}\}_{i=1}^{d_r}$ as searching dimensions and initial values, to find the minimum solution of the $SSD$: $\bar{S}_K^{(N)}$ }
\STATE {\textbf{Output}: $\bar{S}_K^{(N)}$ is the empirical Karcher mean.} 
\end{algorithmic}
\end{algorithm}

\section{Asymptotic Theory}
\label{sec: AsyTheory}


In this section, we will show that the empirical depth converges to the population depth as the sample size goes to infinity for $h$-depth and modified $h$-depth, defined in section \ref{sec: Methods}. Firstly, for the $h$-depth, we have the following conclusion:
\begin{theorem}
Based on Definitions \ref{def:hdep} and \ref{def:hdep_emp}, let $D(\cdot; P_S)$ represent the $h$-depth for a point process on $[0, T]$ in the probability space $(\Omega, \mathcal F, P_S)$ and $\hat{D}(\cdot; \{S_i\}_{i=1}^{N})$ is the corresponding sample version with sample size $N$. Then, for any input $s \in \Omega$, 
\begin{equation}
\hat{D}(s; \{S_i\}_{i=1}^{N}) \to D(s; P_S) \text{ a.s. }
\end{equation}
\end{theorem}
\noindent This theorem implies the convergence of sample $h$-depth to the population $h$-depth, and can be easily proven using the Strong Law of Large Numbers. 

For the modified $h$-depth, the convergence of the depth value will be based on the convergence of the estimated center. As defined in section \ref{center_estimation}, the Karcher mean can be treated as the solution in $\Omega$ that minimizes the average of squared distances, where the distance is $d_{k,2}$ with $\mathbb{L}^2$ norm.  In fact, we can generalize the Karcher mean definition by using any $d_{k,p}$ distance for $p \geq 1$, given in the following form:
\begin{definition} 
Let $S$ be a random point process on $[0, T]$ in the probability space $(\Omega, \mathcal F, P_S)$.  The generalized Karcher mean of $S$ is defined to be 
\begin{equation}
\mu_{K} = \argmin_{t \in \Omega} \mathbb E[d_{K,p}^2(t,S)],
\label{eq:kmp}
\end{equation}
where $d_{K,p}(\cdot, \cdot)$ is the metric with kernel function $K(\cdot;T)$ and $\mathbb{L}^p$ norm where $p\geq 1$. 
\end{definition}

\begin{definition} 
Let $\{S_i\}_{i=1}^{N}$ be a collection of $N$ independent point processes on $[0, T]$. The empirical generalized Karcher mean of this process, based on $\{S_i\}_{i=1}^{N}$, is defined as 
\begin{equation}
\bar{S}_K^{(N)} = \argmin_{t \in \Omega} \frac{1}{N} \sum_{i=1}^{N} d_{K,p}^2(t,S_i)
\label{eq:kmps}
\end{equation}
\end{definition}
\noindent It is apparent that Definitions \ref{def:karcher} and \ref{def:ekarcher} are special cases with $p=2$ in the generalized definitions, respectively. Then the convergence theorem is given as follows:
\begin{theorem}
\label{asy:theory}
$S$ is a random point process on $[0, T]$ in the probability space $(\Omega, \mathcal F, P_S)$, where the number of events $|S|$ has a constant upper bound $D>0$. $\{S_i\}_{i=1}^{N}$ is the set of independent event time vectors from $S$. Then we have:
\begin{enumerate}
\item The minimum average of squared distance in $\{S_i\}_{i=1}^{N}$ will converge to the minimum expected squared distance in $(\Omega, \mathcal F, P_S)$ almost surely, in other words,
\begin{equation}
\min_{t \in \Omega} \frac{1}{N} \sum_{i=1}^{N} d_{K,p}^2(t,S_i) 
\to
\min_{t \in \Omega} \mathbb E[d_{K,p}^2(t,S)]
\quad \text{a.s.}
\label{mineq1}
\end{equation}

\item The empirical generalized Karcher mean converges to the generalized Karcher mean almost surely, in other words, 
\begin{equation}
\bar{S}_K^{(N)} \to \mu_{K} \quad \text{a.s.}
\label{argmineq1}
\end{equation}
The almost-surely convergence from set $\bar{S}_K^{(N)}$ to set $\mu_{K}$ means $\Limsup_{n\to\infty} \bar{S}_K^{(N)} \subset \mu_{K}  \text{ a.s.}$, where $\Limsup_{n\to\infty} \bar{S}_K^{(N)}$ is the Kuratowski upper limit \citep{kuratowski2014topology}:
\begin{equation*}
\Limsup_{n\to\infty}\bar{S}_K^{(N)} = \{ x\in\Omega \quad | \quad \liminf_{N\to\infty} d_{K,p}(x,\bar{S}_K^{(N)})=0  \}
\end{equation*}
where $d_{K,p}(x,\bar{S}_K^{(N)})=\inf\{ d_{K,p}(x,\bar{S}) \quad | \quad \bar{S} \in \bar{S}_K^{(N)} \}$.

\item Let $D(\cdot; s_c)$ be the modified $h$-depth in Definition \ref{def:hdep_cen} with center $s_c$. Suppose the Karcher mean contains a unique element: $\mu_{K}=\{s_c^{(p)}\}$. $\hat{s}_c^{(N)}$ is any element from the empirical Karcher mean $\bar{S}_K^{(N)}$ based on $\{S_i\}_{i=1}^{N}$. Then, for any $s \in \Omega$, 
\begin{equation}
D(s; \hat{s}_c^{(N)}) \to D(s; s_c^{(p)})  \text{ a.s. }
\end{equation}

\end{enumerate}

\noindent Remark 1: In Eqns. \eqref{eq:kmp}, \eqref{eq:kmps}, and \eqref{mineq1}, $\min$ is used instead of $\inf$ because the minimum value can be achieved in both the empirical version and the population version.

\noindent  Remark 2: The theorem will still hold when the square power of the distance in Eqn. \eqref{mineq1} and in the definitions of $\bar{S}_K^{(N)}$ and $\mu_{K}$ are generalized to any $r \geq 1$, i.e. replacing $d_{K,p}^2$ by $d_{K,p}^r$ with any $r \in [1,\infty)$ in the Karcher mean and empirical Karcher mean.
\label{asyT1}
\end{theorem}

\noindent The proof of Theorem \ref{asy:theory}, together with the two remarks, is shown in Appendix M. With the first two points of the theorem, we can confirm that the minimum of the $SSD$ converges to the minimum of the population expected squared distance, irrespective of the optimal solution sets in sample and population. It also indicates the convergence of the estimated center set to its population version. The last point of the theorem suggests that the modified $h$-depth with sample Karcher mean as center converges to the one with population Karcher mean as the center. 

\section{Application Results}
\label{sec: AppResult}
In this section, we will apply the proposed depth framework, $h$-depth and modified $h$-depth, on point process observations in simulations and real experimental data. 

\subsection{Simulation Studies}
\label{App:sim}

\subsubsection{Homogeneous Poisson Process}
\label{sim:hpp}

\begin{figure}[!h]
\begin{center}
\begin{tabular}{cc}
\includegraphics[width=55mm]{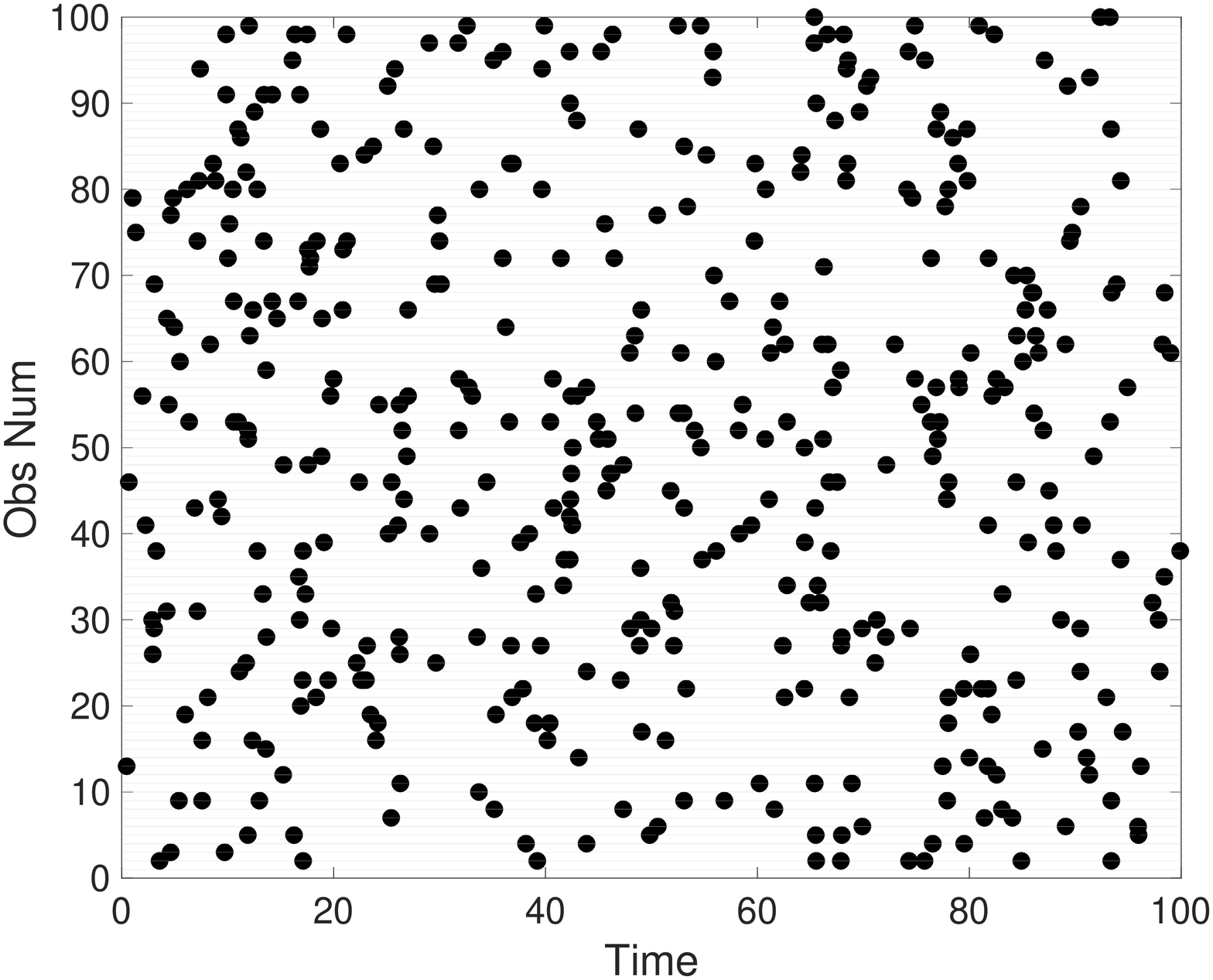} & \includegraphics[width=55mm]{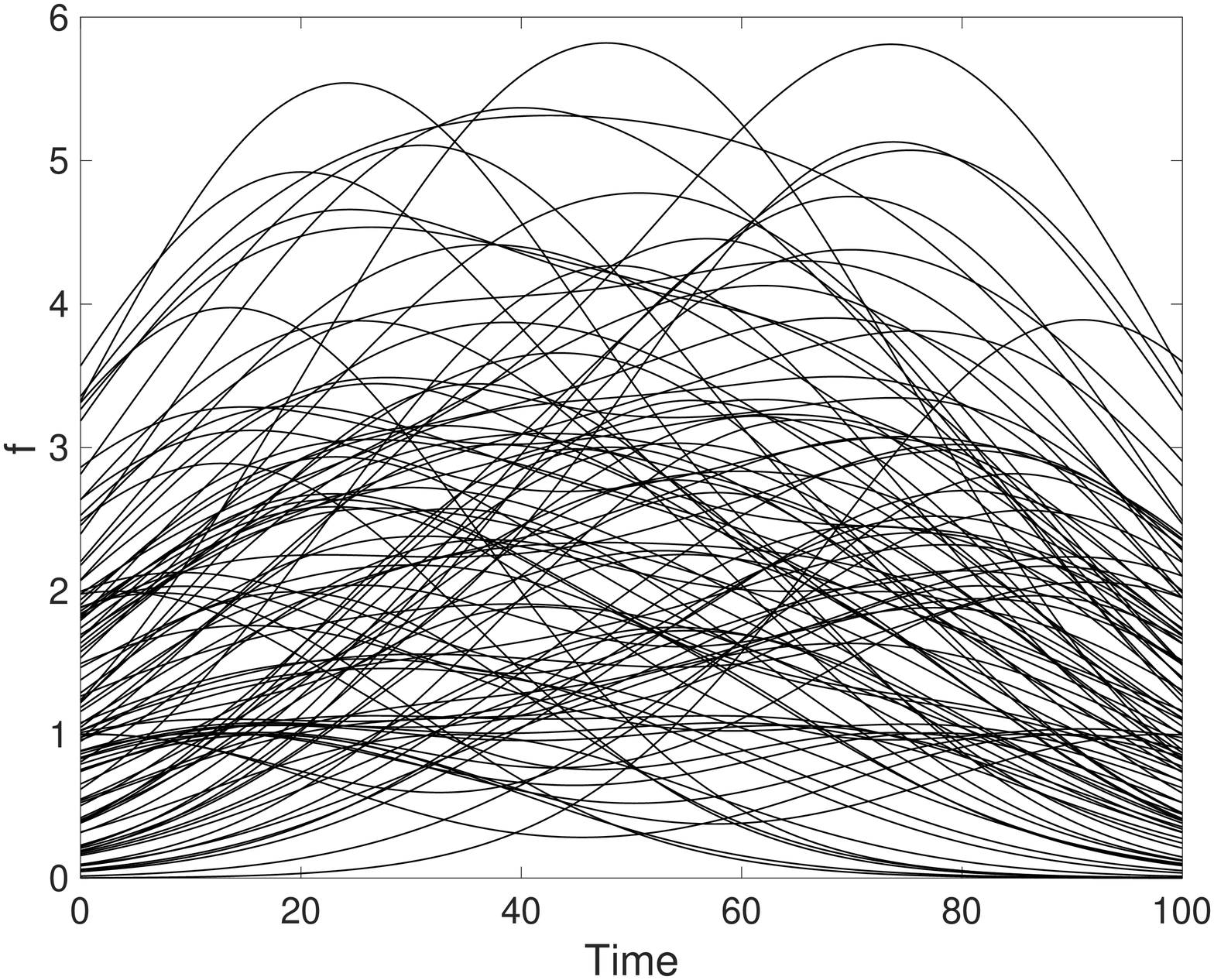} \\
(a) Original processes & (b) Smoothed processes
\end{tabular}
\end{center}
\caption{Simulation on HPP(0.045).  
(a) 100 realizations from HPP(0.045) on interval $[0, 100]$.  Each row represents one realization, where each dot is one event.
(b) Smoothed processes of these 100 realizations using a modified Gaussian kernel. }
\label{fig:hpp_sample}
\end{figure}

We will at first illustrate the depth methods on observations from HPP($\lambda$) in a finite time interval, where $\lambda$ is a constant mean value. 100 independent realizations from HPP(0.045) on $[0, 100]$ are generated, where the raster plot of these processes is shown in Figure \ref{fig:hpp_sample}(a).   Basically, the number of events in each process follows a Poisson distribution with mean 4.5 (varying from 1 to 10 in these 100 realizations), and the event time in every observation is uniformly distributed on [0, 100].  We then smooth the realizations using a kernel in Eqn. \eqref{eq:gk} with $c_1=1, c_2=10$. The smoothed processes are shown in Figure \ref{fig:hpp_sample}(b).  

Based on the $h$-depth and modified $h$-depth in section \ref{sec: Methods}, we compute the sample depth value for each given process where the constant $h$ is set to $T=100$. For the modified $h$-depth, the center is determined in four ways: set by prior information, estimated by RJMCMC annealing, line search and combined method. As the Poisson process is homogeneous, we can intuitively set the center as $(20,40,60,80)$. The output of the center estimation is shown in Table \ref{tab:hchpp}. We can see that all the estimated centers have lower $SSD$ than the intuitive center.  In particular, the combined method has the superior performance. It has the smallest $SSD$ and also the smallest time cost.  We will only use this method for the modified $h$-depth in comparison with other depth methods.

\begin{table}[h]
\centering
\centering
\begin{tabular}{|c|c|c|c|}
\hline
\textbf{Method} & \textbf{Estimated Center} & \textbf{SSD} & \textbf{Time Cost (s)} \\ \hline
Intuitive  & $[20.00,40.00,60.00,80.00]$ & 14217 & $ - $ \\ \hline
RJMCMC Annealing  & $[15.79,33.52,72.28,76.51]$  & 13964 & 256.12\\ \hline
Line Search & $[17.07,34.13,63.84,81.79]$ & 13877 & 207.14\\ \hline
Combined & $[23.02,28.29,67.59,78.88]$ & 13858 & 38.91\\ \hline
\end{tabular}
\caption{Center estimation output for the HPP simulation}
\label{tab:hchpp}
\end{table}


\begin{figure}[!h]
\begin{center}
\begin{tabular}{cc}
\includegraphics[width=55mm]{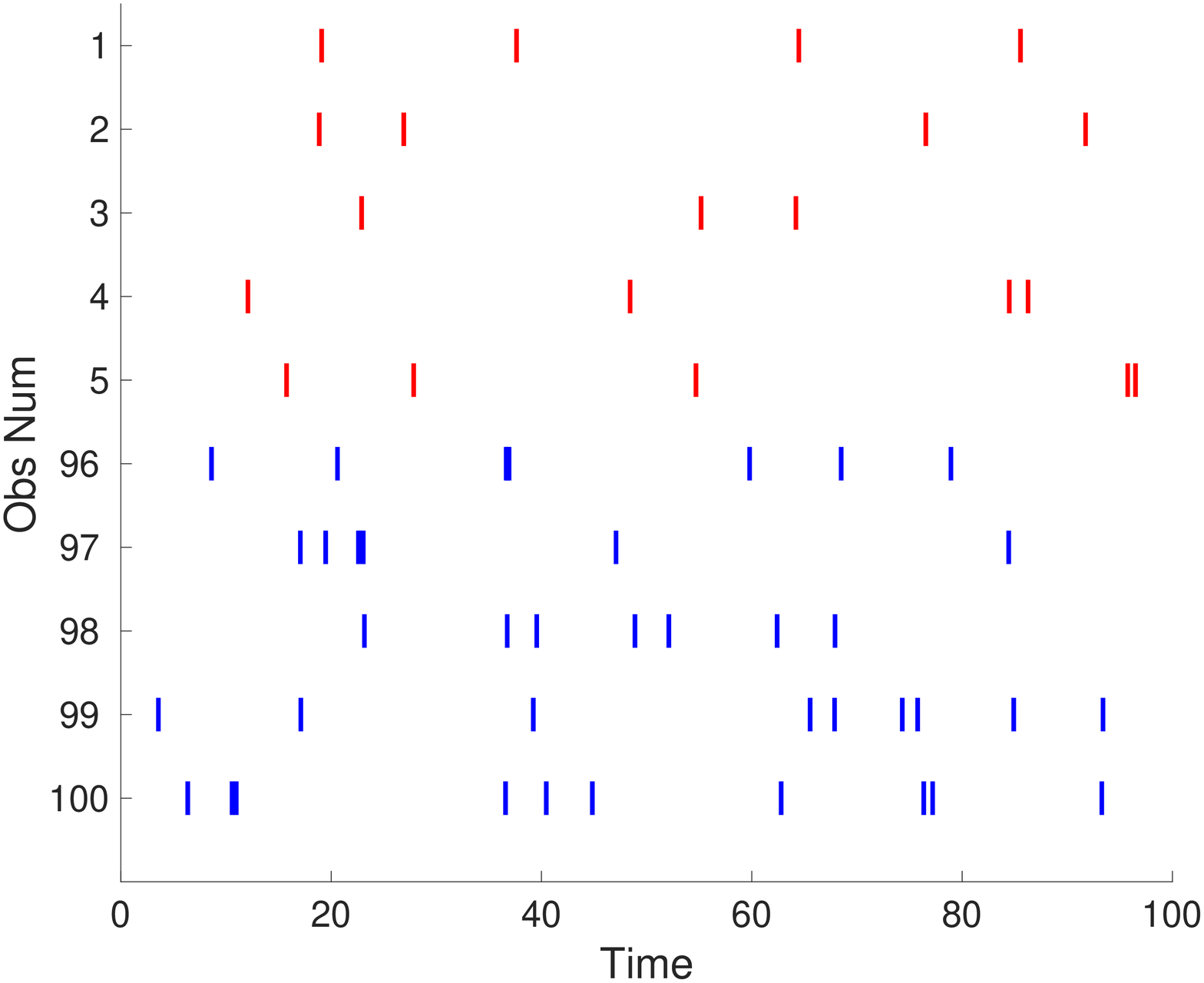} & \includegraphics[width=55mm]{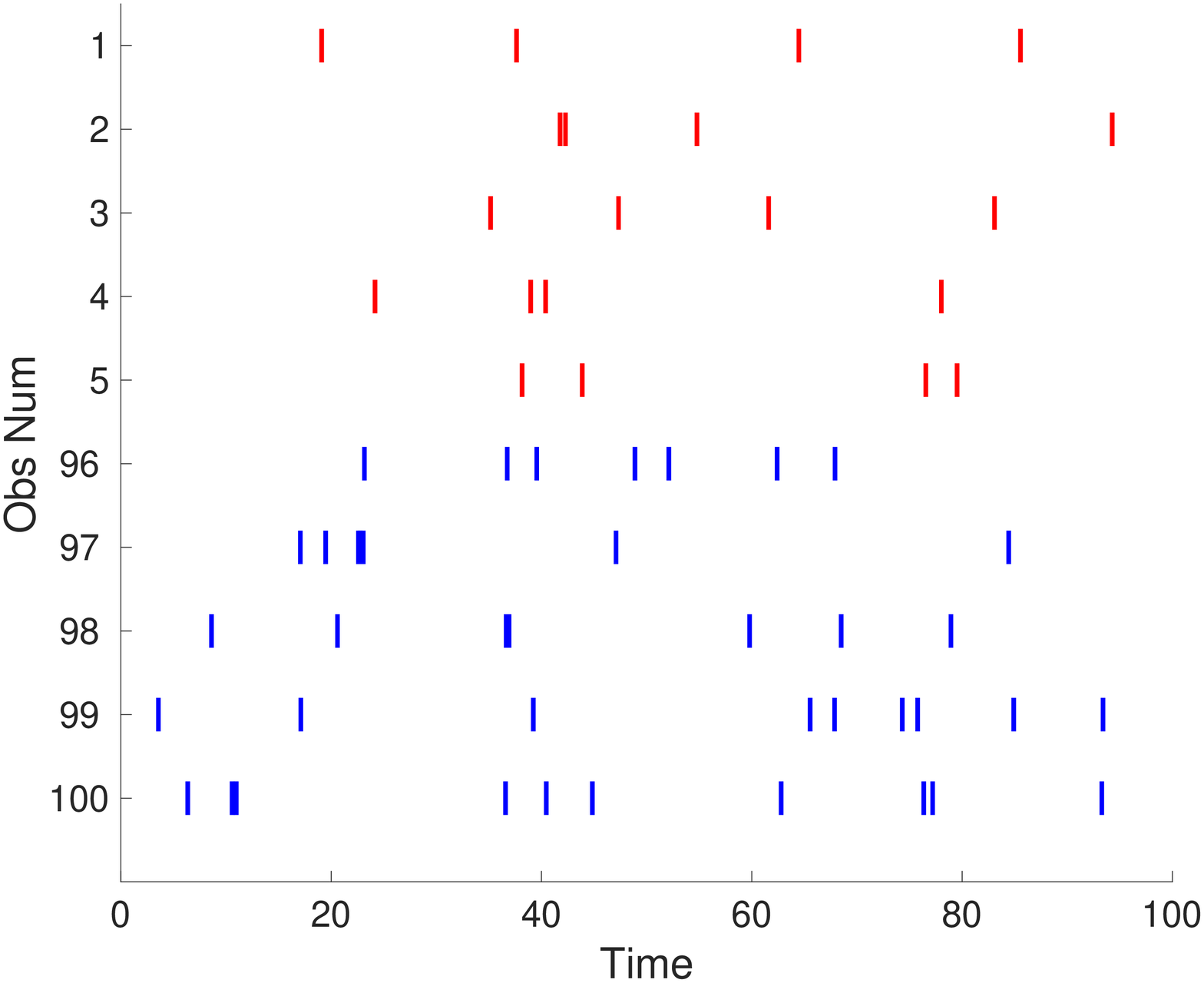} \\
(a) $h$-depth & (b) Modified $h$-depth: intuitive center \\
\includegraphics[width=55mm]{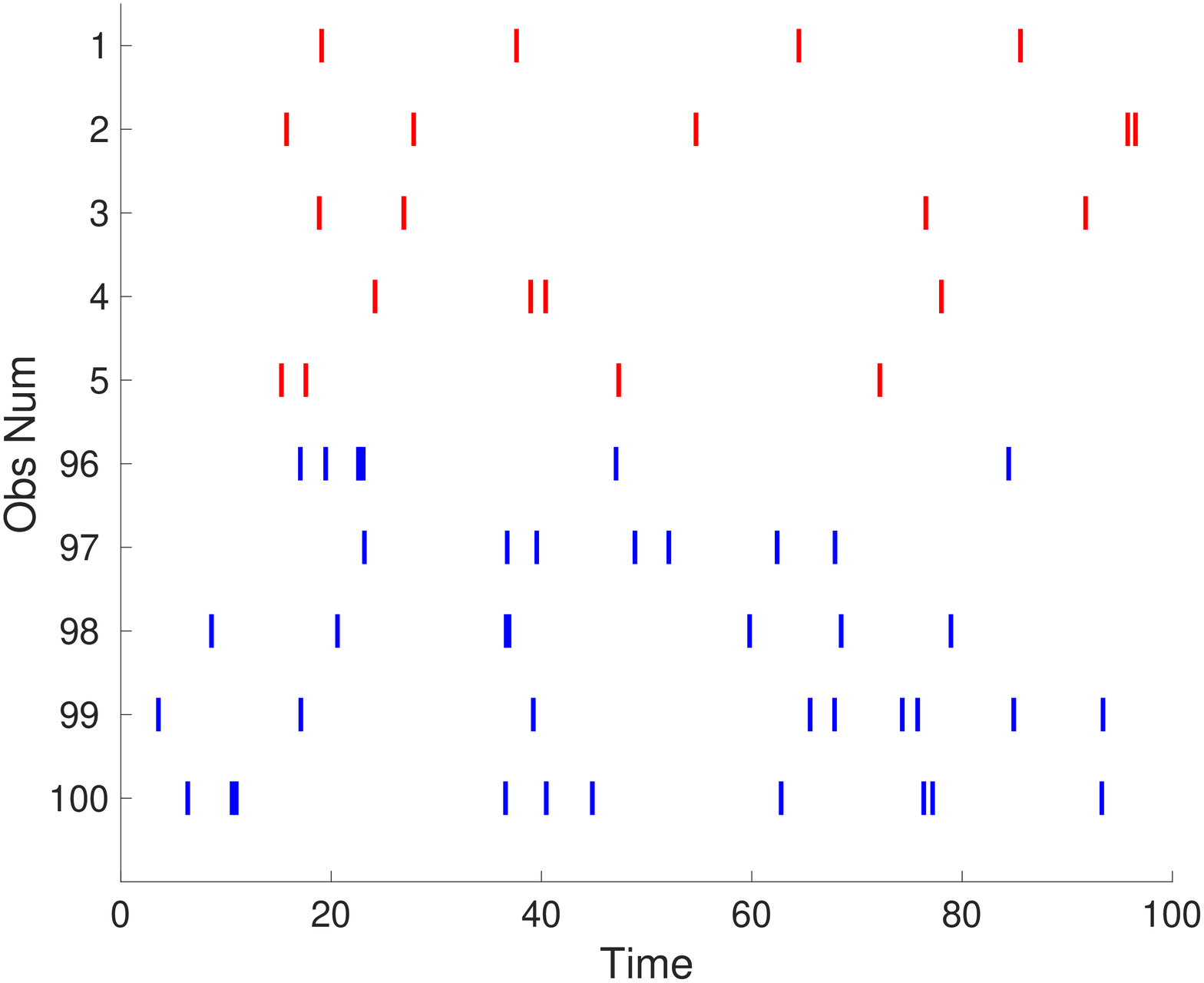} & \includegraphics[width=55mm]{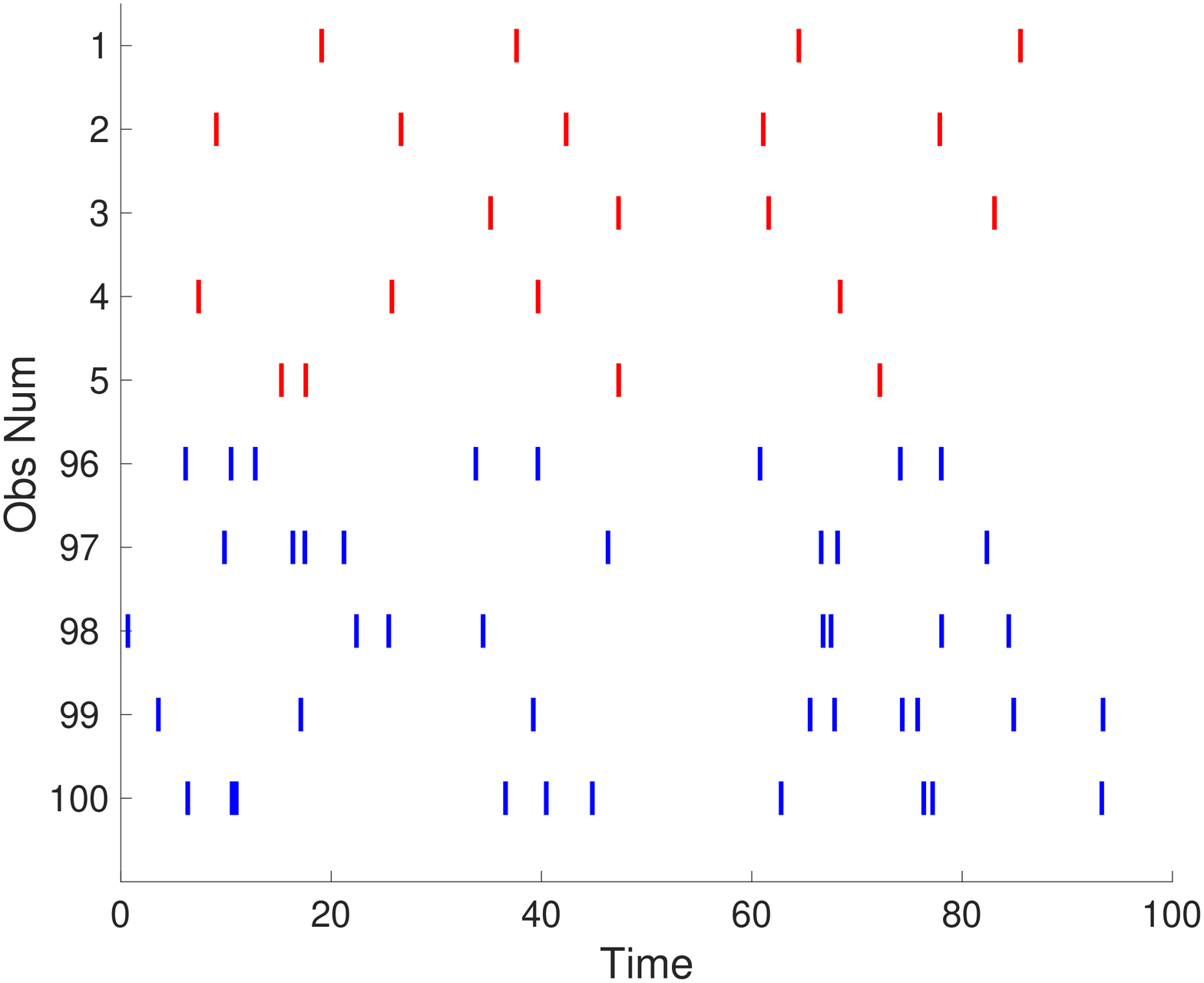}  \\
(c) Modified $h$-depth: estimated center & (d) Mahalanobis depth 
\end{tabular}
\end{center}
\caption{Top 5 (red) and bottom 5 (blue) processes ranked by the depth values in the HPP simulation. 
(a) By the $h$-depth.
(b) By the modified $h$-depth with intuitive center in Table \ref{tab:hchpp}.
(c) By the modified $h$-depth with center estimated using the combined method in Table \ref{tab:hchpp}. 
(d) By the Mahalanobis depth.}
\label{fig:hpp_top5}
\end{figure}

\begin{figure}[!h]
\begin{center}
\begin{tabular}{ccc}
\includegraphics[width=50mm]{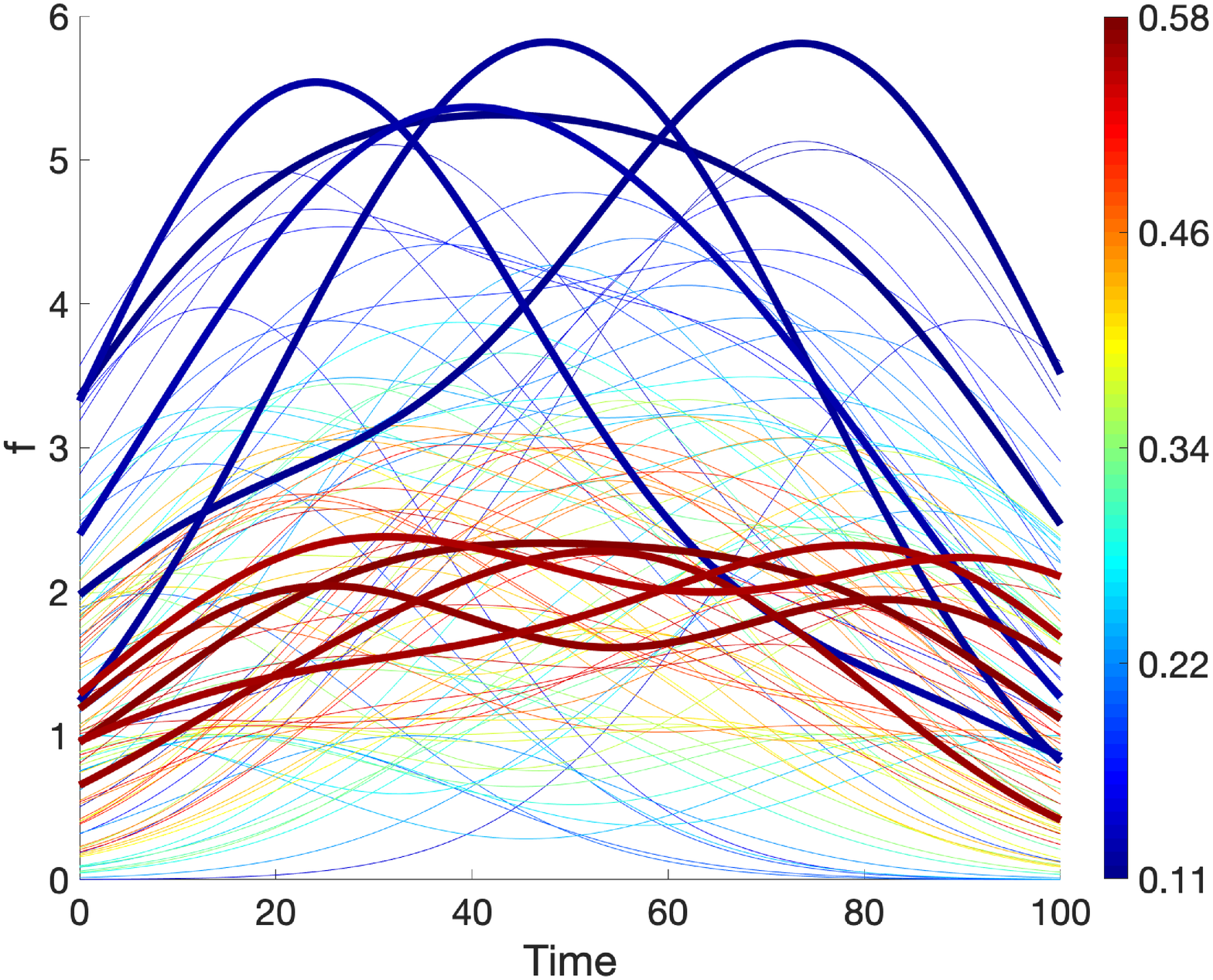} & \includegraphics[width=50mm]{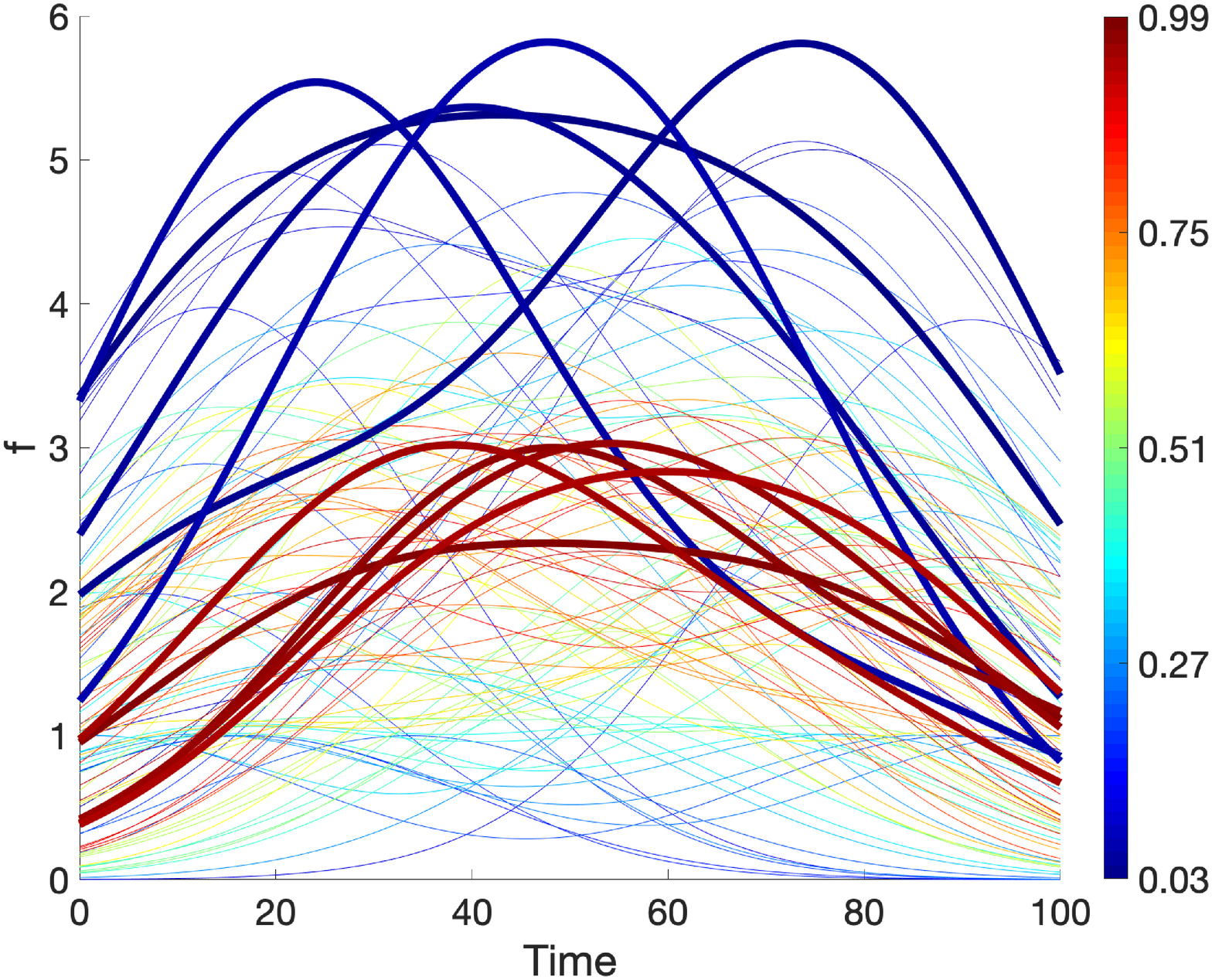} & \includegraphics[width=50mm]{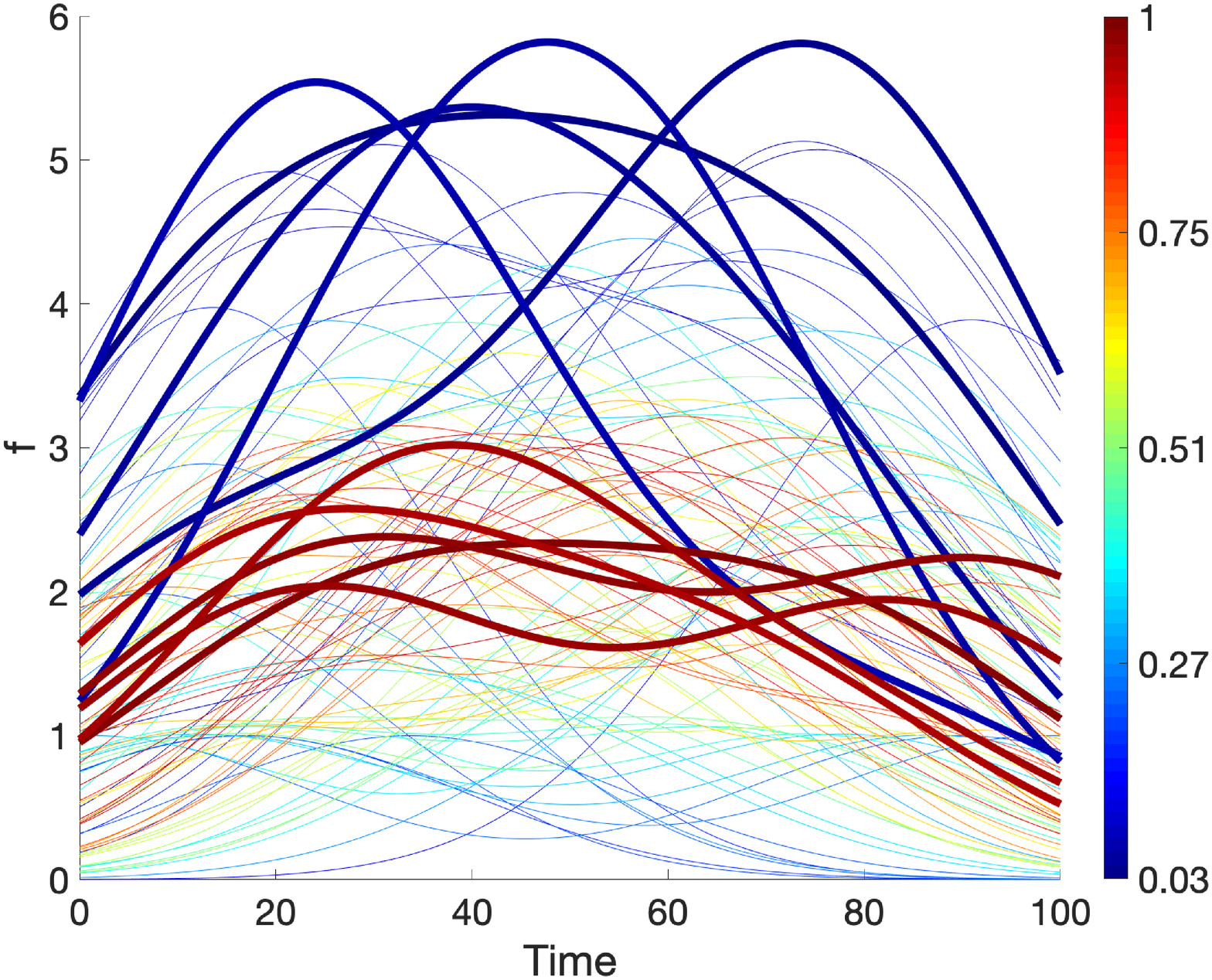}\\
(a) $h$-depth & (b) Modified $h$-depth:  & (c) Modified $h$-depth:  \\
& intuitive center & estimated center
\end{tabular}
\end{center}
\caption{Color-mapped smoothed processes based on depth values for the HPP simulation, where top 5 (red) and bottom 5 (blue) are marked with thick lines.  
(a) By the $h$-depth.
(b) By the modified $h$-depth with the intuitive center.
(c) By the modified $h$-depth with the estimated center using the combined method. }
\label{fig:hpp_heat}
\end{figure}

The depth values for all 100 processes are computed and the observations are then ranked. For better illustration, instead of showing all the ranked processes, we display only the top 5 and bottom 5 in Figure \ref{fig:hpp_top5}. For comparison, we also adopt the generalized Mahalanobis depth for HPP case with the power weight $r=1$ and include the ranking result. It can be seen that the outputs from the $h$-depth and modified $h$-depth are similar to the output from the Mahalanobis depth.  That is, observations with more uniformly distributed events turn to have larger depth values.  In addition, the process with the number of events around the mean 4.5 have larger depth values.

We then display the corresponding smoothed processes in  Figure \ref{fig:hpp_heat}. Note that the generalized Mahalanobis depth is not included as it is not based on smoothing functions. All three plots demonstrate a center-outward decreasing depth value structure that observations whose smoothing curves are closer to the middle of all the curves will have larger depth values. Overall, these simulation results are reasonable and consistent with the basic notion of center-outward ranks. 

\subsubsection{Inhomogeneous Poisson Process}
\label{sim:ipp}

\begin{figure}[!h]
\begin{center}
\begin{tabular}{cc}
\includegraphics[width=55mm]{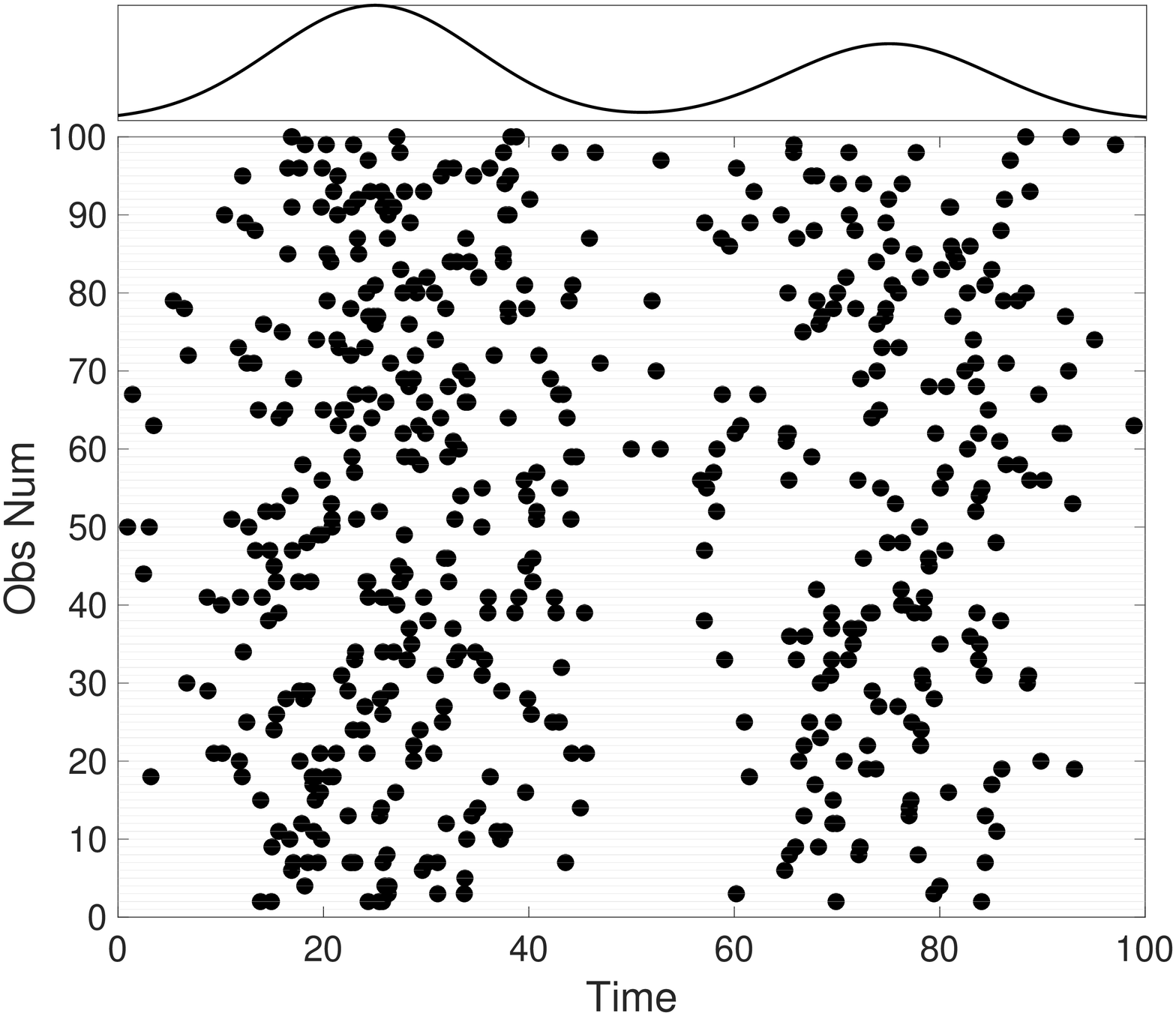} & \includegraphics[width=55mm]{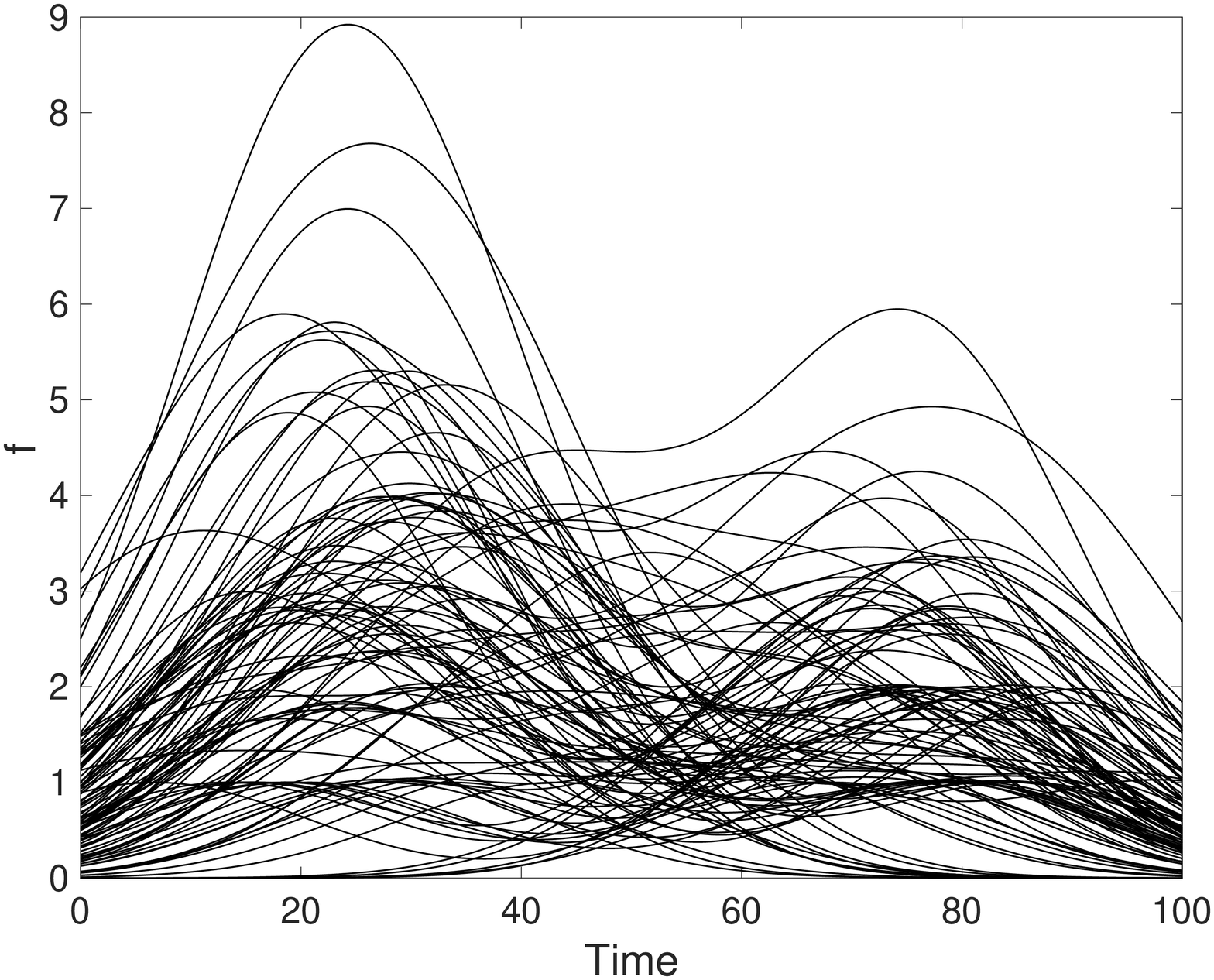} \\
(a) Original processes  & (b) Smoothed processes
\end{tabular}
\end{center}
\caption{An IPP simulation with a dual-peak intensity function. (a) 100 realizations from IPP[$\lambda(t)$] on interval [0,100]. Each row represents one realization, where each dot is one event. The intensity $\lambda(t)$ is shown on the top of all realizations. (b) Smoothed processes of the 100 realizations in (a) using the modified Gaussian kernel. 
}
\label{fig:ipp_sample}
\end{figure}

We will further apply the depth on point process using observations from IPP[$\lambda(t)$] to evaluate the perforamnce. The rate function $\lambda(t)$ is chosen as $\lambda(t)=3\phi(t;25,10)+2\phi(t;75,10)$ and the event time interval is still $[0,100]$, where $\phi(\cdot ;\mu,\sigma)$ is the density for normal distribution with mean $\mu$ and standard deviation $\sigma$. As $\lambda(t)$ contains two peaks, the simulated event times will mainly distribute around 25 and 75. The kernel function is also the modified version of Gaussian kernel with $c_1=1, c_2=25$
. 100 samples are generated. The point process plot and smoothing curve plot are shown in Figure \ref{fig:ipp_sample}.

Similar to section \ref{sim:hpp}, we let the constant $h$ equal $T=100$ in the $h$-depth and modified $h$-depth. Because there is no intuitive center in this case, only the estimated centers are used in the modified $h$-depth. We again use the RJMCMC annealing, line search and combined method. The output is shown in Table \ref{tab:hcipp}. It can be seen that the three estimated centers are very similar. The line search method and the combined method both have the lowest $SSD$, whereas the latter one is much more efficient. That is, the combined method results in the superior performance, and therefore we will only use this method for the modified $h$-depth in comparison with other depth methods. 

%
%

\begin{table}[h]
\centering
\centering
\begin{tabular}{|c|c|c|c|}
\hline
\textbf{Method} & \textbf{Estimated Center} & \textbf{SSD} & \textbf{Time Cost (s)} \\ \hline
RJMCMC Annealing   & $[13.73,27.94,32.69,62.85,79.38]$  & 14152 & 210.61\\ \hline
Line Search (SGD) & $[18.88,24.32,37.21,65.36,82.36]$ & 14063 & 159.05\\ \hline
Combined & $[18.83,24.32,37.22,65.36,82.38]$ & 14063 & 48.77 \\ \hline
\end{tabular}
\caption{Center estimation output for the IPP simulation}
\label{tab:hcipp}
\end{table}

\begin{figure}[!ht]
\begin{center}
\begin{tabular}{ccc}
\includegraphics[width=50mm]{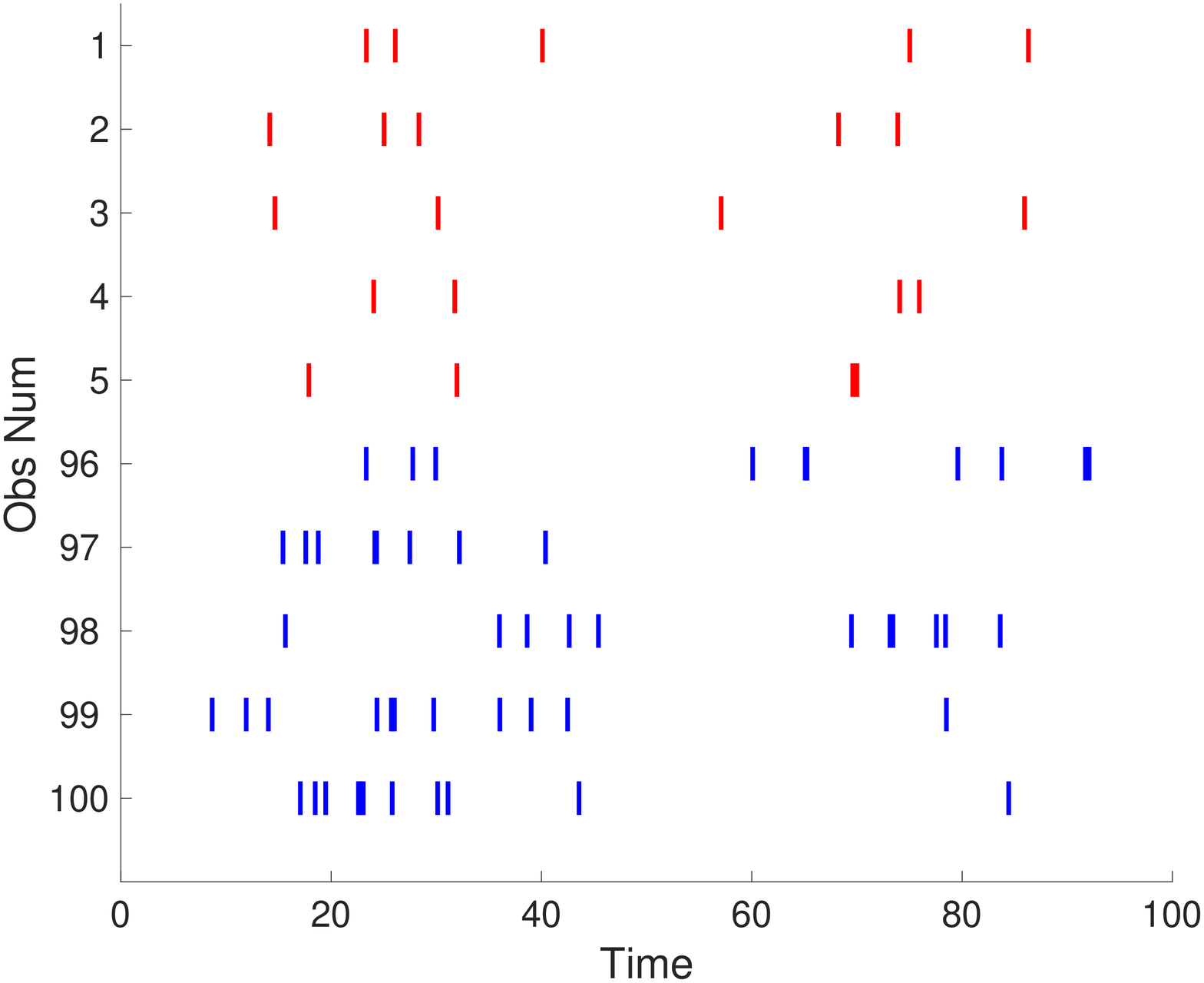} & \includegraphics[width=50mm]{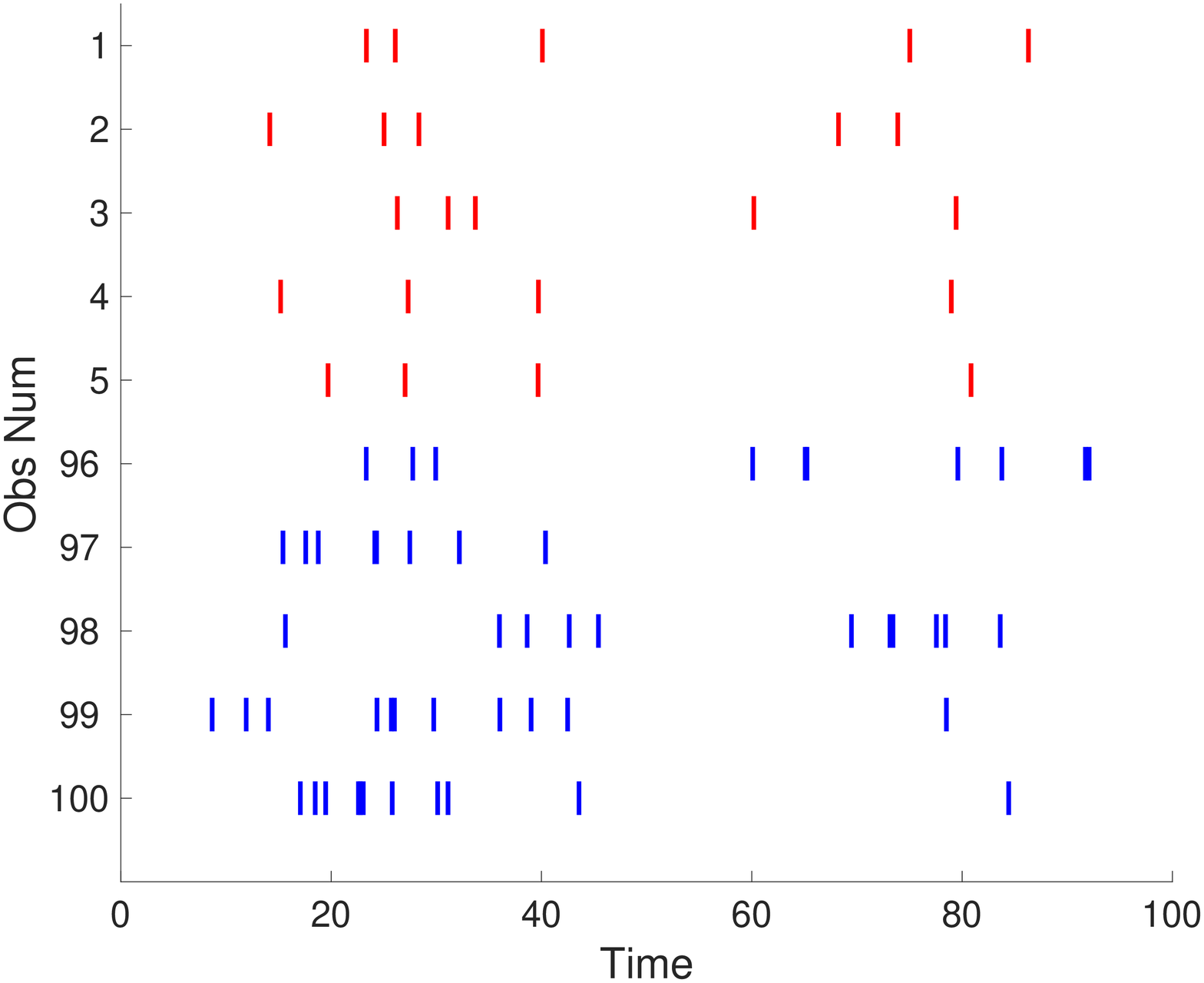} & \includegraphics[width=50mm]{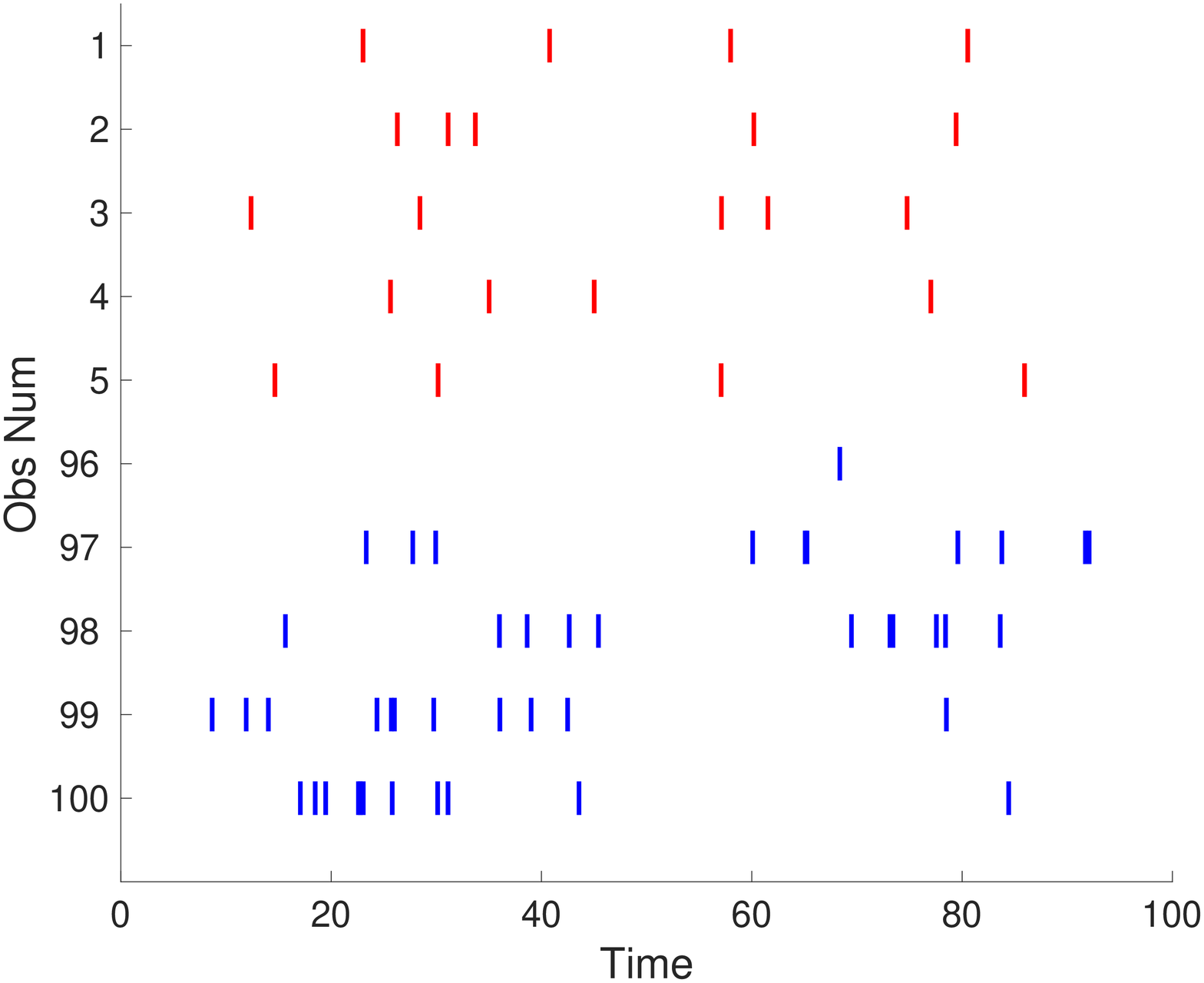} \\
(a) $h$-depth & (b) Modified $h$-depth & (c) Mahalanobis depth
\end{tabular}
\end{center}
\caption{Top 5 (red) and bottom 5 (blue) processes ranked by the depth values for the IPP simulation. 
(a) By the $h$-depth.
(b) By the modified $h$-depth with center estimated using the combined method in Table \ref{tab:hcipp}. 
(c) By the Mahalanobis depth. }
\label{fig:ipp_top5}
\end{figure}

\begin{figure}[!ht]
\begin{center}
\begin{tabular}{cc}
\includegraphics[width=55mm]{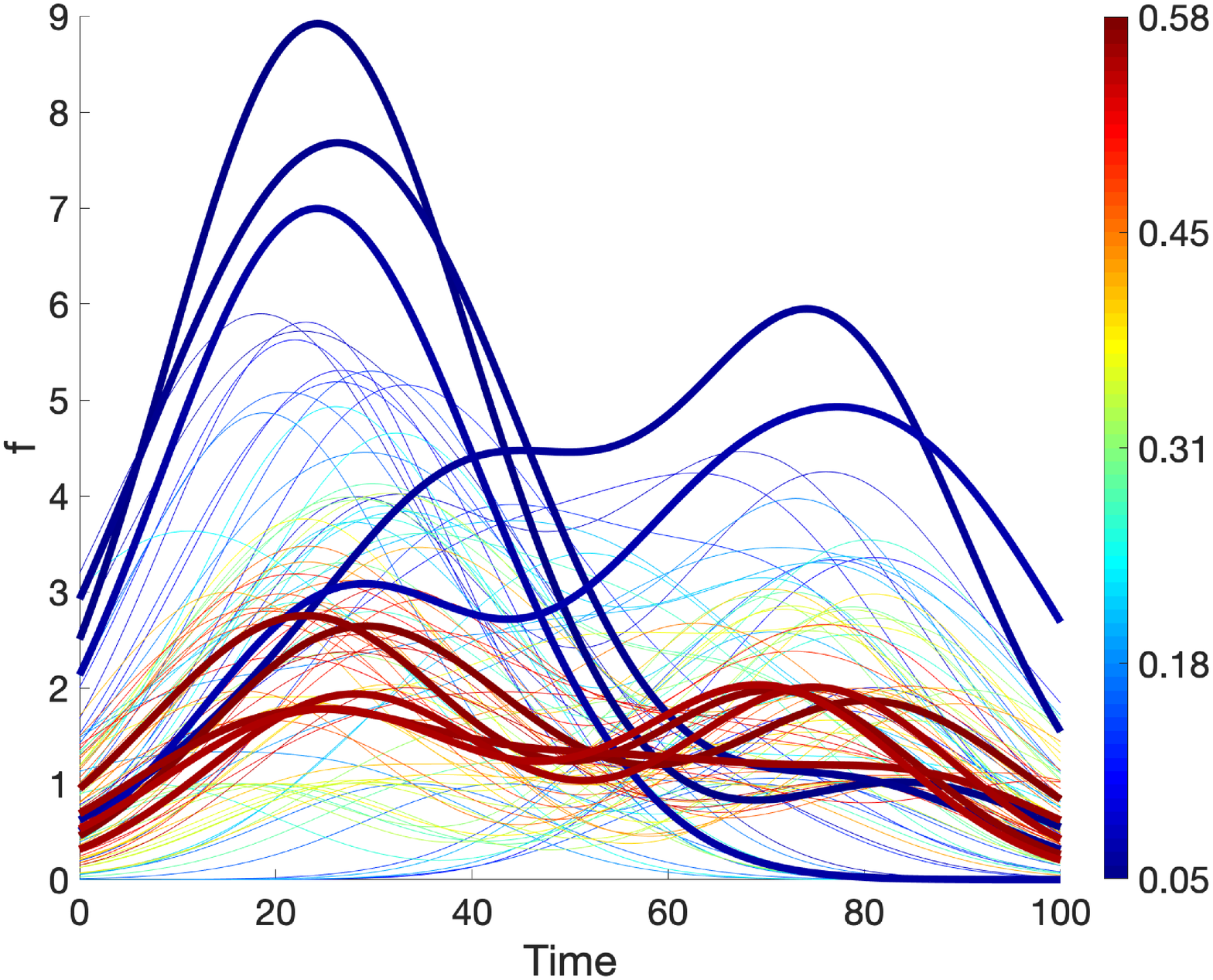} & \includegraphics[width=55mm]{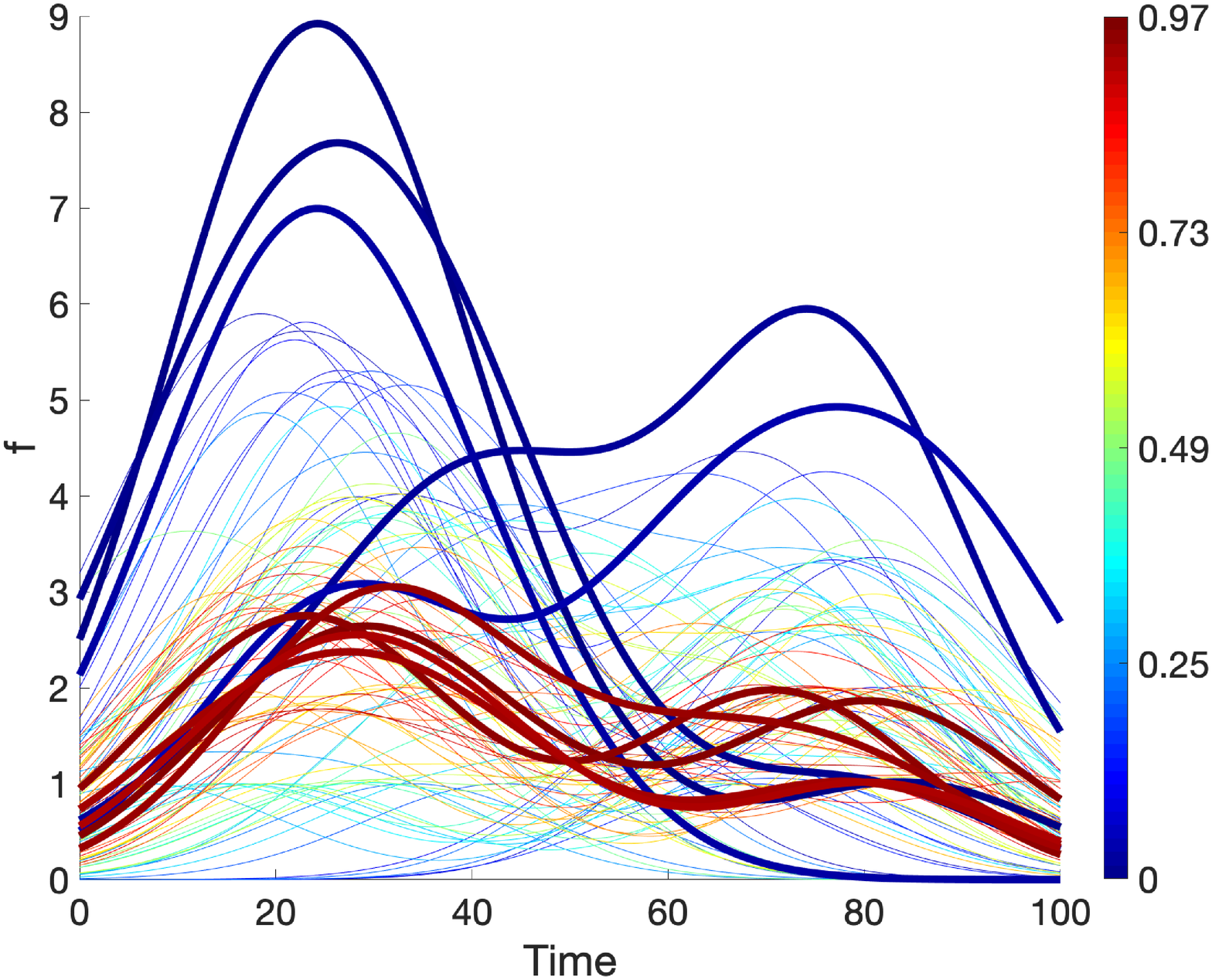} \\
(a) $h$-depth & (b) Modified $h$-depth 
\end{tabular}
\end{center}
\caption{Color-mapped smoothed processes based on depth values for the IPP simulation, where top 5 (red) and bottom 5 (blue) are marked with thick lines.  
(a) By the $h$-depth.
(b) By the modified $h$-depth with center estimated using the combined method. }
\label{fig:ipp_heat}
\end{figure}

After computing the depth for all the observations, we obtain the ranking result. Figure \ref{fig:ipp_top5} shows the top 5 and bottom 5 ranked observations. Similar to the HPP case, we include the result from the generalized Mahalanobis depth with $r=1$ for comparison. It can be seen that the outputs from $h$-depth and modified $h$-depth are similar -- observations with event times around 25 and 75 and number of events around 5 have larger depth. In contrast, for the Mahalanobis depth, we can also observe that processes with number of events around 5 have larger depth. However, rather than concentrating at two peak time locations, the top 5 processes have event time more uniformly distributed. It indicates that $h$-depth and modified $h$-depth can better capture the pattern inside the sample.  If further taking into account the true $\lambda(t)$ that the peak at 25 is higher than the peak at 75, we should expect to see more event times concentrated around 25 than around 75. Comparing Figure \ref{fig:ipp_top5} (a) and (b), we can see that for the modified $h$-depth, all the top-5 observations have more events around 25 than around 75, whereas this is not clearly shown for the $h$-depth. 
The smoothed processes ranked by the $h$-depth and modified $h$-depth are shown in Figure \ref{fig:ipp_heat}. They both exhibit a clear center-outward structure.  That is, observations whose smoothing curves are peaked at around 25 and 75 have relatively larger depth values.

In summary, based on the result from the HPP and IPP simulations, we can see that the proposed $h$-depth and modified $h$-depth both can properly build a center-outward rank on the given point process data.

\subsection{Experimental data application}
\label{App:exp}

We will examine the proposed depth framework  using a real spike train recording, where the data were included in the Quantitative Single-Neuron Modeling Competition 2009 \citep{naud2009quantitative} and were accessible at \url{http://dx.doi.org/10.6080/K0PN93H3}. 
The detailed experiment description can be found in \citep{carandini2007thalamic,sincich2007transmission}. Briefly, the experiment was performed on rhesus monkeys that retinal input (visual stimulus) was applied and extracellular potentials were recorded for both the retinal (pre-synaptic) and the geniculate (post-synaptic) simultaneously. There were 10 seconds stimulus, where the first 5 seconds were the same for all trials and the last 5 seconds were unique for each trial. In total, 76 trials were performed. Since the task in the competition was to predict the post-synaptic spikes given the pre-synaptic spikes, only 38 pairs of (post-synaptic, pre-synaptic) were fully given. 
We will use these spike train data to test the smoothing depth framework through a classification task: 
In the training set, we have labeled observations for pre-synaptic (abbreviated as ``pre-group'') and post-synaptic (abbreviated as ``post-group'').
In the testing set, we compute the depth for each observation in the two groups. The group with a larger depth value will be the predicted label.

\begin{figure}[!h]
\begin{center}
\begin{tabular}{cc}
\includegraphics[width=100mm]{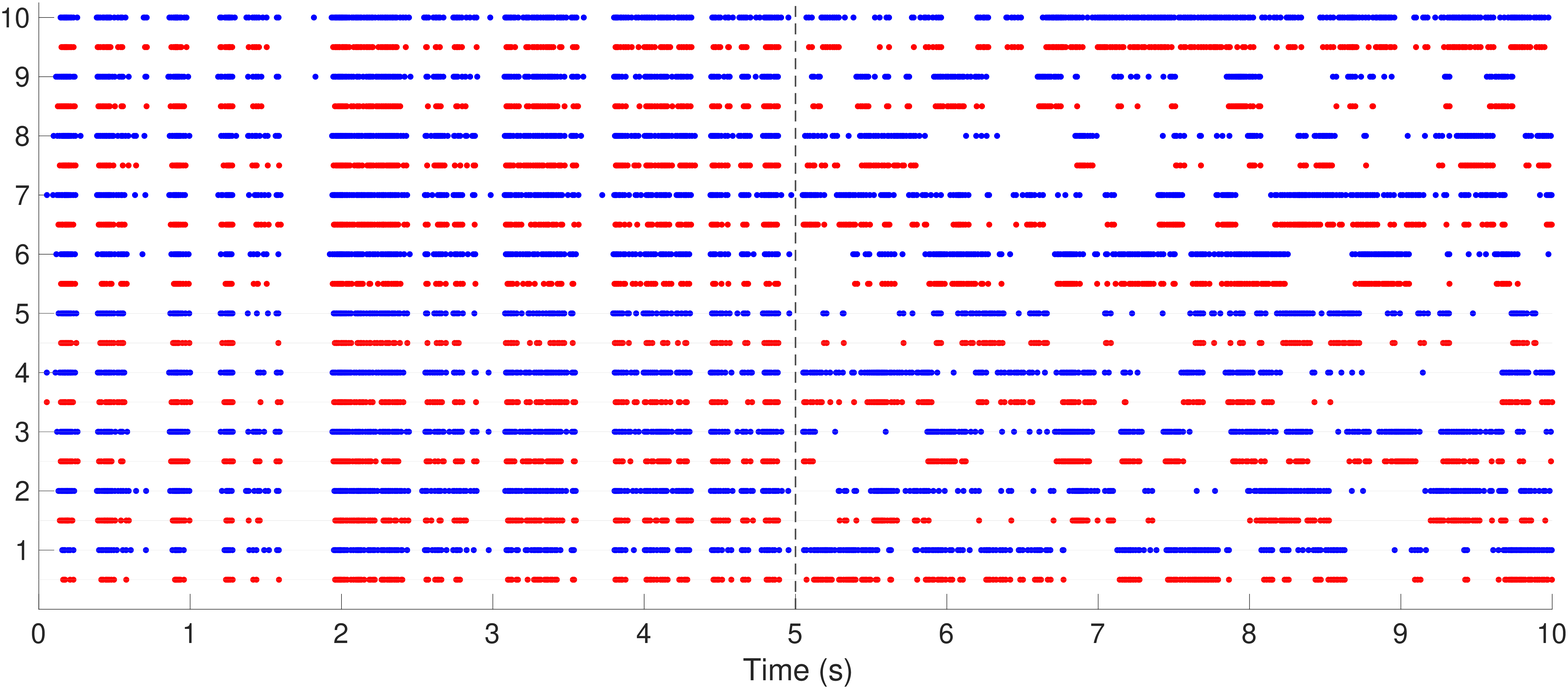}
\end{tabular}
\end{center}
\caption{The point process plot for the spike train observations. Blue: pre-group; Red: post-group. 10 example pairs were selected and displayed. }
\label{fig:exp_data1}
\end{figure}

Before the classification, an exploratory data analysis is performed. The spike times are within the 10 seconds range and the point process plot is shown in Figure \ref{fig:exp_data1}. 10 example pairs of pre-group, post-group observations are shown. As stated in the experiment description, the stimuli were the same for the first 5 seconds and different for the last 5 seconds.  This well explains why the signals have very similar pattern in the first 5 seconds.  Therefore, it is reasonable to separate the trial into two parts, first 5 seconds and last 5 seconds, and we can smooth them in different ways. 

\begin{figure}[!h]
\begin{center}
\begin{tabular}{cc}
\includegraphics[width=100mm]{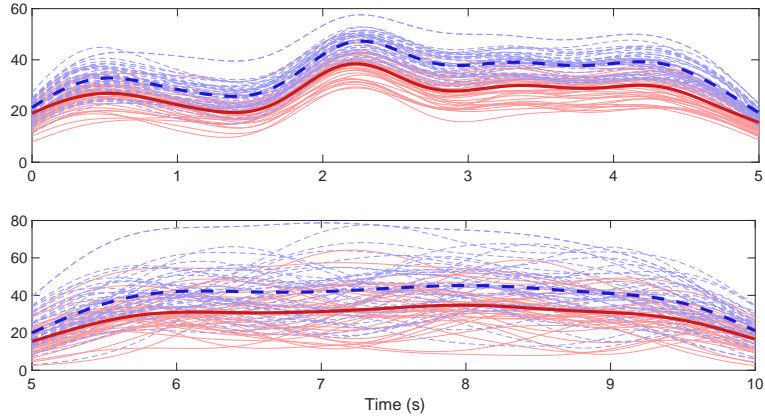}
\end{tabular}
\end{center}
\caption{Smoothed observations for pre-group and post-group in the first 5 seconds (top panel) and last (bottom panel) 5 seconds. Blue dashed lines: pre-group; Red solid lines: post-group. The thick lines denote the estimated Karcher means in the two groups, repsectively.} 
\label{fig:exp_data2}
\end{figure}


After separation, we apply the modified Gaussian kernel to smooth the observations with $c_1=1$, $c_2=100$ for first 5 seconds and $c_1=1$, $c_2=50$ for last 5 seconds. The smoothing curve plot is given in Figure \ref{fig:exp_data2}. Using the combined method in section \ref{center_estimation}, we estimate the Karcher mean for the two groups and the corresponding smoothing curves are shown as thick lines. It can be seen that the Karcher mean curves are able to catch the shape patterns in the first 5 seconds as the observations are nearly aligned, while in the last 5 seconds, the Karcher mean is nearly flat since the observations are very noisy. 

The classification is done by a 4-fold cross validation. 
We compute the accuracy and F1 score on the predicted labels versus the true labels to measure the classification performance. In addition to the classical $h$-depth and modified $h$-depth with center given, we conduct the classification through the generalized Mahalanobis depth on point process \citep{liu2017generalized},  as well as the band depth and modified band depth \citep{lopez2009concept} on the smoothed observations. 

\begin{table}[ht]
\centering

\begin{tabular}{l|c|c|c|c|c|c|}
\hline
\multicolumn{1}{|c|}{\multirow{2}{*}{\textbf{Method}}} & \multicolumn{2}{c|}{\textbf{Accuracy}}                                & \multicolumn{2}{c|}{\textbf{F1 score (pre)}}                          & \multicolumn{2}{c|}{\textbf{F1 score (post)}}                          \\ \cline{2-7} 
\multicolumn{1}{|c|}{}                        & first 5s                      & last 5s                       & first 5s                      & last 5s                       & first 5s                      & last 5s                       \\ \hline                             
\multicolumn{1}{|l|}{Classical $h$-depth}   & 78.95\%            & 65.79\%           & 79.49\%               & 61.76\%              & 78.38\%                & 69.05\%              \\ \hline
\multicolumn{1}{|l|}{Modified $h$-depth}    & \textbf{81.58\%}            & \textbf{76.32\%}           & 81.08\%               & \textbf{75.68\%}              & \textbf{82.05\%}                & \textbf{76.92\%}              \\ \hline
\multicolumn{1}{|l|}{Band depth}         & 78.95\%            & 61.84\%           & 77.14\%               & 53.97\%              & 80.49\%                & 67.42\%              \\ \hline
\multicolumn{1}{|l|}{Modified band depth} & \textbf{81.58\%}            & 71.05\%           & \textbf{81.58\%}               & 72.50\%              & 81.58\%                & 69.44\%              \\ \hline
\multicolumn{1}{|l|}{Mahalanobis depth} &  \textbf{81.58\%}          & 73.68\%           &  \textbf{81.58\%}             & 72.97\%              & 81.58\%                & 74.36\%              \\ \hline
\end{tabular}

\caption{The classification result.  F1 score (pre) and F1 score (post) are computed according to the recall and precision on pre-group classification and post-group classification. The boldface indicates best results in each column.  }
\label{tab:exp_data}
\end{table}

The classification result is shown in Table \ref{tab:exp_data}.  All methods are able to provide reasonable classification power on the spike train data. The overall results on the first 5 seconds is better than the last 5 seconds.
For the last 5 seconds, it can be seen that the modified $h$-depth has the best accuracy (76.32\%) and F1 scores (75.68\%, 76.92\%). For the first 5 seconds, the modified $h$-depth, modified band depth and Mahalanobis depth all have the best accuracy (81.58\%), while the modified $h$-depth is better on the post group (F1 score 82.05\%) and the other two are better on the pre-group (F1 score 81.58\%). In summary, we can see that the smoothing depth framework provides relatively accurate classification in the practical point process observations.

\section{Summary and Future Work}
\label{sec: SumFW}

In this paper, we at first proposed a kernel smoothing representation for point process observations. 
We found that the smoothing procedure builds a bijective mapping between them.  We then defined a proper metric distance bewteen the smoothing curves and transformed the problems from point processes to smooth functions. 
Based on the notion of $h$-depth on functions, we proposed two methods to form the depth structure on point process observations, i.e. the classical $h$-depth and the modified $h$-depth with center given. 
We then proposed a center-estimation method for the modified $h$-depth, with the idea of minimizing the Karcher mean through a combination of the RJMCMC annealing and line search.
During the tests on the simulated data, both classical $h$-depth and modified $h$-depth resulted in reasonable outcomes: a center-outward decreasing depth structure that observations with smoothing curves closed to the center will have higher depth. 
In the real neuronal spike train data, we tested the new depth structures with a depth-based classification task and both methods resulted in accurate classifications. 

Our investigation is only the starting point of the smoothing curve exploration and there is much to be further studied in the future. We have proposed the modified Gaussian kernel that satisfies the four basic requirements. 
We will explore other kernel functions that also satisfy the requirements. In the modified $h$-depth, a combination of RJMCMC annealing and line search has been given as the method to estimate the Karcher mean, which serves as the ``center''. There should be other ways to define the ``center'' and other estimation methods.  These are all very good topics to further develop depth function on point process. Finally,  more functional depth methods, other than $h$-depth, can be adopted to smoothed point processes, such as the band depth and modified band depth. Their ranking performance needs thorough investigations. 

\bibliographystyle{agsm}
\bibliography{main}

@inproceedings{tukey1975mathematics,
  title={Mathematics and the picturing of data},
  author={Tukey, John W},
  booktitle={Proceedings of the International Congress of Mathematicians, Vancouver, 1975},
  volume={2},
  pages={523--531},
  year={1975}
}

@article{liu1990notion,
  title={On a notion of data depth based on random simplices},
  author={Liu, Regina Y and others},
  journal={The Annals of Statistics},
  volume={18},
  number={1},
  pages={405--414},
  year={1990},
  publisher={Institute of Mathematical Statistics}
}

@article{lopez2009concept,
  title={On the concept of depth for functional data},
  author={L{\'o}pez-Pintado, Sara and Romo, Juan},
  journal={Journal of the American Statistical Association},
  volume={104},
  number={486},
  pages={718--734},
  year={2009},
  publisher={Taylor \& Francis}
}

@article{liu2017generalized,
  title={Generalized Mahalanobis depth in point process and its application in neural coding},
  author={Liu, Shuyi and Wu, Wei and others},
  journal={The Annals of Applied Statistics},
  volume={11},
  number={2},
  pages={992--1010},
  year={2017},
  publisher={Institute of Mathematical Statistics}
}

@article{nieto2016topologically,
  title={A topologically valid definition of depth for functional data},
  author={Nieto-Reyes, Alicia and Battey, Heather and others},
  journal={Statistical Science},
  volume={31},
  number={1},
  pages={61--79},
  year={2016},
  publisher={Institute of Mathematical Statistics}
}

@article{grove1973conjugatec,
  title={How to conjugatec 1-close group actions},
  author={Grove, Karsten and Karcher, Hermann},
  journal={Mathematische Zeitschrift},
  volume={132},
  number={1},
  pages={11--20},
  year={1973},
  publisher={Springer}
}

@article{green1995reversible,
  title={Reversible jump Markov chain Monte Carlo computation and Bayesian model determination},
  author={Green, Peter J},
  journal={Biometrika},
  volume={82},
  number={4},
  pages={711--732},
  year={1995},
  publisher={Oxford University Press}
}

@article{metropolis1953equation,
  title={Equation of state calculations by fast computing machines},
  author={Metropolis, Nicholas and Rosenbluth, Arianna W and Rosenbluth, Marshall N and Teller, Augusta H and Teller, Edward},
  journal={The journal of chemical physics},
  volume={21},
  number={6},
  pages={1087--1092},
  year={1953},
  publisher={American Institute of Physics}
}

@incollection{van1987simulated,
  title={Simulated annealing},
  author={Van Laarhoven, Peter JM and Aarts, Emile HL},
  booktitle={Simulated annealing: Theory and applications},
  pages={7--15},
  year={1987},
  publisher={Springer}
}

@article{geman1984stochastic,
  title={Stochastic relaxation, Gibbs distributions, and the Bayesian restoration of images},
  author={Geman, Stuart and Geman, Donald},
  journal={IEEE Transactions on pattern analysis and machine intelligence},
  number={6},
  pages={721--741},
  year={1984},
  publisher={IEEE}
}

@book{kuratowski2014topology,
  title={Topology},
  author={Kuratowski, Kazimierz},
  volume={1},
  year={2014},
  publisher={Elsevier}
}

@article{barnett1976ordering,
  title={The ordering of multivariate data},
  author={Barnett, Vic},
  journal={Journal of the Royal Statistical Society: Series A (General)},
  volume={139},
  number={3},
  pages={318--344},
  year={1976},
  publisher={Wiley Online Library}
}

@article{oja1983descriptive,
  title={Descriptive statistics for multivariate distributions},
  author={Oja, Hannu},
  journal={Statistics \& Probability Letters},
  volume={1},
  number={6},
  pages={327--332},
  year={1983},
  publisher={Elsevier}
}

@article{liu1993quality,
  title={A quality index based on data depth and multivariate rank tests},
  author={Liu, Regina Y and Singh, Kesar},
  journal={Journal of the American Statistical Association},
  volume={88},
  number={421},
  pages={252--260},
  year={1993},
  publisher={Taylor \& Francis Group}
}

@article{zuo2000general,
  title={General notions of statistical depth function},
  author={Zuo, Yijun and Serfling, Robert},
  journal={Annals of statistics},
  pages={461--482},
  year={2000},
  publisher={JSTOR}
}

@article{cuevas2007robust,
  title={Robust estimation and classification for functional data via projection-based depth notions},
  author={Cuevas, Antonio and Febrero, Manuel and Fraiman, Ricardo},
  journal={Computational Statistics},
  volume={22},
  number={3},
  pages={481--496},
  year={2007},
  publisher={Springer}
}

@article{cuesta2008random,
  title={The random Tukey depth},
  author={Cuesta-Albertos, Juan Antonio and Nieto-Reyes, Alicia},
  journal={Computational Statistics \& Data Analysis},
  volume={52},
  number={11},
  pages={4979--4988},
  year={2008},
  publisher={Elsevier}
}

@book{wand1994kernel,
  title={Kernel smoothing},
  author={Wand, Matt P and Jones, M Chris},
  year={1994},
  publisher={CRC press}
}

@article{carandini2007thalamic,
  title={Thalamic filtering of retinal spike trains by postsynaptic summation},
  author={Carandini, Matteo and Horton, Jonathan C and Sincich, Lawrence C},
  journal={Journal of vision},
  volume={7},
  number={14},
  pages={20--20},
  year={2007},
  publisher={The Association for Research in Vision and Ophthalmology}
}

@article{sincich2007transmission,
  title={Transmission of spike trains at the retinogeniculate synapse},
  author={Sincich, Lawrence C and Adams, Daniel L and Economides, John R and Horton, Jonathan C},
  journal={Journal of Neuroscience},
  volume={27},
  number={10},
  pages={2683--2692},
  year={2007},
  publisher={Soc Neuroscience}
}

@inproceedings{naud2009quantitative,
  title={Quantitative single-neuron modeling: competition 2009},
  author={Naud, Richard and Berger, Thomas and Bathellier, Brice and Carandini, Matteo and Gerstner, Wulfram},
  booktitle={Front. Neur. Conference Abstract: Neuroinformatics 2009},
  pages={1--8},
  year={2009}
}

\end{document}